\documentclass[twocolumn,superscriptaddress,showpacs,prd,
aps,amsmath,amssymb,nofootinbib]{revtex4}
\usepackage{graphicx}
\usepackage{amssymb}

\usepackage{color}

\newcommand{\hn}{{\mathfrak n}}
\newcommand{\hB}{{\mathfrak B}}

\newcommand{\Madm}{M}
\newcommand{\MK}{M_{\rm K}}
\newcommand{\beq}{\begin{equation}} 
\newcommand{\eeq}{\end{equation}} 
\newcommand{\beqn}{\begin{eqnarray}} 
\newcommand{\eeqn}{\end{eqnarray}} 
\newcommand{\pa}{\partial}
\newcommand{\na}{\nabla}

\newcommand{\gabu}{g^{\alpha\beta}}
\newcommand{\gabd}{g_{\alpha\beta}}

\newcommand{\gmabd}{\gamma_{ab}}

\newcommand{\albe}{{\alpha\beta}}
\newcommand{\beal}{{\beta\alpha}}
\newcommand{\albega}{{\alpha\beta\gamma}}

\newcommand{\qabu}{q^{\alpha\beta}}

\newcommand{\Tabu}{T^{\alpha\beta}}
\newcommand{\Tab}{T^\alpha{}\!_\beta}
\newcommand{\Tba}{T_\alpha{}^\beta}
\newcommand{\TabFd}{T^{\rm F}_{\alpha\beta}}
\newcommand{\TabFu}{T_{\rm F}^{\alpha\beta}}
\newcommand{\TabF}{T_{{\rm F}}{}\!^\alpha{}\!_\beta}

\newcommand{\TabMu}{T_{\rm F}^{\alpha\beta}}
\newcommand{\TabM}{T_{{\rm F}}{}\!^\alpha{}\!_\beta}
\newcommand{\Gab}{G^\alpha{}\!_\beta}

\newcommand{\Gabu}{G^{\alpha\beta}}
\newcommand{\Rab}{R^\alpha{}\!_\beta}
\newcommand{\Rabd}{R_{\alpha\beta}}

\newcommand{\Fabu}{F^{\alpha\beta}}
\newcommand{\Fabd}{F_{\alpha\beta}}

\newcommand{\Fcdu}{F^{\gamma\delta}}
\newcommand{\Fcdd}{F_{\gamma\delta}}
\newcommand{\CF}{C_{\rm F}}
\newcommand{\oabd}{\omega_{\alpha\beta}}

\newcommand{\PhiE}{{\Phi^{\rm E}}}
\newcommand{\PhiEi}{{\Phi^{\rm E}_i}}
\newcommand{\QE}{Q^{\rm E}}
\newcommand{\QEi}{Q^{\rm E}_i}
\newcommand{\QEm}{Q^{\rm E}_{\rm m}}
\newcommand{\Hori}{{\cal H}^{\pm}}
\newcommand{\Horiplus}{{\cal H}^{+}}

\newcommand{\zD}{{\raise1.0ex\hbox{${}^{\ \circ}$}}\!\!\!\!\!D}
\newcommand{\alone}{{\raise0.5ex\hbox{${}^{\ 1}$}}\!\!\!\!\alpha}
\newcommand{\Od}{{O}}

\newcommand{\dSab}{dS_{\alpha\beta}}
\newcommand{\dSa}{dS_{\alpha}}

\newcommand{\QL}{Q_{L}}

\newcommand{\dl}{\delta}

\newcommand{\Dlgab}{\Delta g_{\alpha\beta}}

\newcommand{\Dl}{\Delta}

\newcommand{\Lie}{\mbox{\pounds}}
\newcommand{\Lag}{{\cal L}}
\newcommand{\LagG}{{\cal L}_{\rm G}}
\newcommand{\Lagm}{{\cal L}_{\rm m}}
\newcommand{\LagM}{{\cal L}_{\rm F}}
\newcommand{\LagF}{{\cal L}_{\rm F}}
\newcommand{\LagI}{{\cal L}_{\rm I}}
\newcommand{\Thau}{{\Theta}^\alpha}
\newcommand{\ThGau}{{\Theta}^\alpha_{\rm G}}
\newcommand{\Thmau}{{\Theta}^\alpha_{\rm m}}
\newcommand{\ThMau}{{\Theta}^\alpha_{\rm F}}
\newcommand{\ThFau}{{\Theta}^\alpha_{\rm F}}
\newcommand{\ThFbu}{{\Theta}^\beta_{\rm F}}
\newcommand{\ThIau}{{\Theta}^\alpha_{\rm I}}

\newcommand{\nalam}{\mathrel{\raise0.9ex\hbox{$^\lambda$}\mkern-14mu
\lower0.0ex\hbox{$\nabla$}}}

\newcommand{\dis}{\displaystyle}

\newcommand{\zeroD}{{\raise1.0ex\hbox{${}^{\ \circ}$}}\!\!\!\!\!D}
\newcommand{\zLap}{{\raise1.0ex\hbox{${}^{\ \circ}$}}\!\!\!\!\Delta}
\newcommand{\zna}{\stackrel{{\circ}}{\nabla}}
\newcommand{\zS}{{\raise1.0ex\hbox{${}^{\ \circ}$}}\!\!\!\!\!S}

\newcommand{\hq}{{\hat{q}}}

\begin{document}

\title{
Thermodynamics of magnetized binary compact objects 
} 

\author{K\=oji Ury\=u}
\affiliation{
Department of Physics, University of the Ryukyus, Senbaru, 
Nishihara, Okinawa 903-0213, Japan}
\author{Eric Gourgoulhon}
\affiliation{
Laboratoire Univers et Th\'eories, UMR 8102 du CNRS,
Observatoire de Paris, Universit\'e Paris Diderot, F-92190 Meudon, France}
\author{Charalampos Markakis} 
\affiliation{
Department of Physics, University of Wisconsin-Milwaukee, P.O. Box 413,  
Milwaukee, WI 53201}
\date{28 August 2010}  

\begin{abstract} 

Binary systems of compact objects with electromagnetic field are modeled 
by helically symmetric Einstein-Maxwell spacetimes with 
charged and magnetized perfect fluids.  
Previously derived thermodynamic laws for helically-symmetric perfect-fluid 
spacetimes are extended to include the electromagnetic fields, and electric 
currents and charges; the first law is written as a relation between the change in the 
asymptotic Noether charge $\dl Q$ and the changes in the area and electric 
charge of black holes, and in the vorticity, baryon rest mass, entropy, charge 
and magnetic flux of the magnetized fluid.  
Using the conservation laws of the circulation of magnetized flow 
found by Bekenstein and Oron for the ideal magnetohydrodynamic (MHD) fluid, 
and also for the flow with zero conducting current, 
we show that, for nearby equilibria that conserve the quantities mentioned above, 
the relation $\dl Q=0$ is satisfied.  We also discuss a formulation for computing numerical 
solutions of magnetized binary compact objects in equilibrium 
with emphasis on a first integral of the ideal MHD-Euler equation.  
\end{abstract} 
\pacs{}

\maketitle

\section{Introduction}

Recent observations of anomalous X-ray pulsars, or soft $\gamma$-ray 
repeaters suggest the existence of neutron stars associated with 
magnetic fields strong enough to affect their structures in hydrostationary
equilibrium (see, e.g. \cite{2006csxs.book..547W}).  
Such objects have not been found in binary neutron star systems, 
but hypothetically strongly magnetized neutron stars or black holes 
may form binary neutron star or black hole - neutron star systems.  
In this article, we model such magnetized binary compact objects 
in close circular orbits, assuming that the spacetime and magnetic fields 
satisfy a helical symmetry and that the stars are in equilibrium.  

The helically symmetric spacetime was introduced by 
Blackburn and Detweiler \cite{BD92D94} to model 
binary compact objects in close circular orbits 
in general relativity.  
In such spacetimes, equal amounts of 
ingoing and outgoing radiation are propagating, 
and hence these spacetimes do not admit flat 
asymptotics, because the steady radiation field 
carries an infinite amount of energy.  
Nevertheless, it is expected that such a spacetime has an approximate 
asymptotic region up to a certain radius, where gravitational waves 
are propagating in a curved background, and the energy of 
radiation does not dominate in the gravitational mass of the system.  
Such a solution, however, has not yet been 
calculated successfully in the regime of strong gravity.  
Analogously to Schild's result in electromagnetism 
for two oppositely charged point particles \cite{schild}, 
circular orbits of two point particles have been obtained 
in post-Minkowskian spacetimes \cite{FU06GU07}.
More studies for the helically symmetric spacetimes have been 
reported by several authors 
\cite{BonazGM97,Asad98,Friedman:2001pf,GGB02,Klein:2004be,Torre,
Bruneton:2006ft,BBS,PSWconsortium,YBRUF06}.

In \cite{Friedman:2001pf} (hereafter FUS), 
thermodynamic laws for helically symmetric 
perfect fluid spacetimes have been derived.  
In the first part of this paper, we extend the results of FUS to 
the magnetized perfect-fluid Einstein-Maxwell spacetimes 
with helical symmetry.  
As in FUS, we use a helical Killing vector $k^\alpha$ 
to define a conserved Noether current and associated 
Noether charge $Q$ \cite{Lee:1990nz,wald,iyerwald,iyer,ss77,sorkin91,brown}.
With an appropriate choice of the current and a constant of 
the electric potential, the charge $Q$ becomes finite 
and is independent of the 2-surface $S$ on which it is 
evaluated as long as the matter and black holes 
are enclosed in $S$.   
We obtain the first law, which relates the change 
$\dl Q$ to the changes in the baryon mass, entropy, 
circulation and electric current of the fluid, and 
in the area and electric charge of the black holes.  
Its expression corresponds to the mass 
variation formula for stationary axisymmetric 
spacetimes derived by Carter \cite{Carter73,Carter79} 
(Eq.~(\ref{eq:1stlaw_org}) below
\footnote{The first law Eq.~(\ref{eq:1stlaw_org}) 
is for generic flows that respect the helical symmetry.}).  
Concrete calculations for the variation, $\dl Q$, 
associated with the classical action for 
an Einstein-Maxwell theory coupled with a perfect fluid 
carrying an electric current, 
\beq
\Lag \,=\,
\Big(\,\frac1{16\pi} R \,-\, \epsilon 
\,-\,\frac1{16\pi} \Fabd \Fabu \,+\, A_\alpha j^\alpha
\,\Big)\sqrt{-g}, 
\label{eq:Lag_org}
\eeq
are summarized in Appendices \ref{s:Variation_Lagrangian} and 
\ref{secApp:dlQ} to clarify notation and conventions.

When the late stages of binary 
inspiral are modeled using a sequence of equilibrium 
solutions of helically symmetric perfect fluid spacetimes
(without electromagnetic fields), 
the baryon mass, entropy, and 
circulation of the flow, and the area of each black 
hole are assumed to be held constant (see e.g. 
\cite{BNSCF,UryuCFWL,BHNS_QE,GGB02,BBH_QE}).  
Then, the expression of the first law becomes 
$\dl Q = 0$, or for asymptotically flat systems such as 
the post-Newtonian, or the spatially conformally flat systems, 
$\dl M = \Omega\dl J$, 
as a result of the conservations of those quantities (FUS).  
When electromagnetic fields and electric currents are 
present in neutron stars, the circulation of magnetized flow 
is not conserved in general.  Hence, 
it is not possible to find a sequence of solutions 
along which the first law is simplified as above 
without further assumptions for the flow.  
In other words, in order to approximate binary inspiral just before 
a merger by a sequence of quasi-equilibrium solutions,
one needs to introduce a model for the evolution of 
neutron star spins.  However, as shown in Sec.~\ref{sec:conserv}, 
with an electric current introduced by Bekenstein 
and Oron for a class of ideal magnetohydrodynamic (MHD) flows 
(\cite{OronBB,Bekenstein:2000sf,BB06}, see also \cite{TarapovGorskii} 
for non-relativistic magnetized flow), 
a generalized circulation of magnetized flow is found to be conserved.  
Applying this law -- the generalized Kelvin theorem for ideal MHD -- 
we show that the relation $\dl Q=0$ is satisfied along a sequence of helically 
symmetric equilibria of magnetized binary systems, and that the relation 
$\dl M = \Omega \dl J$ holds for asymptotically flat systems.

The above first law can be applied to actual sequences of solutions,  
and hence in the second part of the paper, in Sec.~\ref{sec:integrability} and 
\ref{secApp:magfluid}, formulations for computing such equilibrium 
solutions of magnetized binary compact objects are discussed.  
In particular, we discuss the first integral of the MHD-Euler equation, 
which is a key to compute equilibria of neutron stars numerically.  
Bekenstein and Oron \cite{Bekenstein:2000sf} have found 
a first integral of the relativistic MHD-Euler equation using 
the same current for the case with ideal MHD irrotational flow, 
and also for the case with the purely convection current.  
As irrotational flow is considered to be more realistic in the 
final inspiral stage of the binary neutron stars and the black hole - 
neutron star binaries \cite{KBC92}, 
we introduce the first integral by Bekenstein and Oron for 
ideal MHD irrotational flow, then derive a somewhat different 
first integral, which may be valid only on an initial hypersurface $\Sigma_t$, 
and write down a set of equations for 
the magnetized irrotational flow suitable for numerical 
computations of binary neutron stars and 
black hole-neutron star binaries in equilibrium.

We follow the conventions and notation in FUS.  
For a one-form $w_\alpha$, the exterior 
derivative $(dw)_\albe$ (within index notation) is defined by 
\beq
(dw)_\albe := \na_\alpha w_\beta - \na_\beta w_\alpha, 
\eeq
and for a two-form $w_{\albe}= w_{[\albe]}$ by 
\beq
(dw)_\albega := 3\,\na_{[\alpha} w_{\beta\gamma]} 
= \na_\alpha w_{\beta\gamma}
+ \na_\beta w_{\gamma\alpha} 
+ \na_\gamma  w_{\alpha\beta}.
\eeq

\section{Thermodynamic laws for 
Einstein-Maxwell spacetime
with charged and magnetized perfect fluid 
}

\subsection{Zeroth law and constancy of the electric potential 
on the Killing horizon}

We consider a globally hyperbolic spacetime $({\cal M}, \gabd)$ 
and a vector field $k^\alpha$ transverse to each 
Cauchy surface (but not necessarily everywhere timelike).
This vector generates the one-parameter family of 
diffeomorphisms $\chi_t$.  The action of $\chi_t$ 
to a spacelike sphere $\cal S$ on a Cauchy surface 
generates a timelike surface, 
${\cal T}({\cal S}) = \cup_t \chi_t({\cal S})$, called 
the \emph{history of} $\cal S$.  
Then, as in FUS, $k^\alpha$ is called a \emph{helical vector} if 
there is a smallest $T>0$ for which $P$ and $\chi_T(P)$ are 
timelike separated for every point $P$ outside of 
the history ${\cal T}({\cal S})$.  
Very often, $k^\alpha$ can be written $k^\alpha = t^\alpha + \Omega \phi^\alpha$,
where $\Omega = 2\pi/T$,  
$t^\alpha$ is a timelike vector 
and $\phi^\alpha$ a spacelike vector that has circular orbits 
with a parameter length $2\pi$ (see, FUS).  

Each Cauchy surface of the helically symmetric 
spacetime does not admit flat asymptotics 
because the energy of the radiation generated 
by a binary equilibrium eventually dominates 
and causes a divergence.  
Therefore, as discussed in FUS, the future (past) 
horizon $\Hori$ in helically symmetric spacetime 
is defined by the boundary of the future (past) domain of 
outer communication ${\cal D}^\pm$, where $P\in \cal M$
is in ${\cal D}^\pm$ if the future (past) timelike curve 
$c(\lambda)$ through $P(:=c(0))$ remains outside of 
${\cal T}({\cal S})$ of each sphere $\cal S$ for a 
certain $\lambda_0$, $\lambda > \lambda_0$.  
It is also shown that, if the history 
${\cal T}({\cal S})$ of a sphere $\cal S$ is in 
${\cal D}^\pm$, the future (past) horizon agrees 
with the chronological past (future) of the history 
${\cal T}$, $\Hori = \pa I^\mp({\cal T})$.  

The conditions of the theorems by Friedrich, R\'acz, and Wald \cite{FRW} 
are modified to make them suitable for helically symmetric spacetimes.
With the assumption that the \emph{null energy condition} 
holds: $R_\albe l^\alpha l^\beta \ge 0$
for any null vector $l^\alpha$,
those theorems yield the following properties:
$\Hori$ are Killing horizons, 
the shear $\sigma_\albe$ and the expansion $\theta$ of 
a null congruence vanish on $\Hori$, the Killing vector 
$k^\alpha$ is parallel to the null generators of the horizons, 
and the surface gravity $\kappa$ of each disconnected horizon 
defined by 
\beq
k^\beta \na_\beta k^\alpha \,=\, \kappa k^\alpha 
\eeq
is constant on each connected component of $\Hori$ (FUS).  

The Raychaudhuri equation, 
\beq
\frac{d\theta}{d\lambda}
\,=\,-\Rabd l^\alpha l^\beta
- 2\sigma_\albe \sigma^\albe
-\frac12 \theta^2,
\label{eq:raychou}
\eeq
is used to demonstrate the above properties. It implies 
$\Rabd l^\alpha l^\beta =0 $ on the Killing horizons 
$\Hori$.  Assuming there exists no material flow through 
the horizon but there exists an electromagnetic field 
$
\Fabd := (dA)_\albe = \na_\alpha A_\beta - \na_\beta A_\alpha,
$
where $A_\alpha$ is the electromagnetic potential one-form,
we have 
\beqn
\Rabd k^\alpha k^\beta 
&=& 8\pi \TabFd k^\alpha k^\beta 
\,=\, 2 F_{\alpha\gamma}F_\beta{}^\gamma k^\alpha k^\beta 
\nonumber\\
&=& \frac12\left(E_\alpha E^\alpha + B_\alpha B^\alpha \right)
\,=\,0 \label{eq:E2_B2_zero}
\eeqn
on $\Hori$, 
where $\TabFd$ is the stress-energy tensor for 
the electromagnetic field, and 
$E_\alpha$ and $B_\alpha$ are the electric and magnetic 
components with respect to the helical vector defined by
\footnote{If $k^\alpha$ would be normalized
by $k_\alpha k^\alpha = -1$, $E_\alpha$ and $B_\alpha$ could be interpreted physically as 
the electric and magnetic fields measured by the observer of 4-velocity $k^\alpha$. 
Note however that in general $k_\alpha k^\alpha \not= -1$; even $k_\alpha k^\alpha =0$ on 
$\Hori$.}
\beq
E_\alpha\,:=\,\Fabd k^\beta, \quad 
B_\alpha\,:=\,\frac12 \epsilon_{\albe\gamma\delta} F^{\beta\gamma}k^\delta .
\label{eq:defEB}
\eeq
Note that, as a consequence of (\ref{eq:E2_B2_zero}), $E^\alpha$ and $B^\alpha$
are both null on $\Hori$. 
Using the Cartan identity, 
\beq
k^\beta (dA)_{\beta\alpha} 
\,=\, \Lie_k A_\alpha \,-\, \na_\alpha (k^\beta A_\beta),
\label{eq:Cartan}
\eeq
and assuming that $A_\alpha$ respects the symmetry $\Lie_k A_\alpha=0$, 
one can introduce an electric potential in the rotating frame 
$E_\alpha \,=\, -\na_\alpha \PhiE$. 
\footnote{
One can avoid the assumption that the field $A_\alpha$ respects 
the helical symmetry.  
Eq.~(\ref{eq:defEB}) implies 
$(dE)_\albe = -k^\gamma (dF)_{\gamma\alpha\beta} -\Lie_k \Fabd = 0$
for $(dF)_\albega=0$ and the symmetry $\Lie_k F_\albe = 0$.  
Hence, from the Poincar\'e lemma, 
${}^\exists \PhiE$ such that $E_\alpha =-\na_\alpha \PhiE$ 
if the domain is connected and simply connected.  
}
Since $E_\alpha k^\alpha = B_\alpha k^\alpha = 0$ and $E_\alpha$ and $B_\alpha$ are both null
on $\Hori$, 
$E_\alpha$ and $B_\alpha$ 
are necessarily parallel to the null generator on $\Hori$.
Then, for any 
vector $\eta^\alpha$ tangent to $\Hori$, 
$\eta^\alpha E_\alpha = - \eta^\alpha \na_\alpha \PhiE= 0$, 
which implies that $\PhiE$ is constant on the Killing horizon 
$\Hori$ \cite{Carter73,Carter79}.

The potential $\PhiE$ is defined globally 
if the domain of outer communications is simply connected, and 
$\PhiE$ is unique up to the constant of integration.  
The constant may be chosen $\PhiE\rightarrow 0$ as 
$r \rightarrow \infty$ for asymptotically flat systems.  
For the helically symmetric system, we set the constant by 
the condition
\beq
\frac1{4\pi}\oint_S k^\gamma A_\gamma \Fabu dS_\albe
\,=\,0
\label{eq:epotconst}
\eeq
on the boundary sphere $S$ which encloses all black holes and 
neutron stars, and on which a family of Noether 
charges is defined in the next section
\footnote{
For an asymptotically flat spacetime, 
the Noether charge defined on $S$ 
with the choice of Eq.~(\ref{eq:epotconst}), 
and then the radius of $S$ taken to be $r \rightarrow \infty$, 
agrees with a choice 
$\PhiE\rightarrow 0$ at $r \rightarrow \infty$
(see, Sec.~\ref{sec:1stlaw}).}.  
The total electric charge of the system is defined by 
the surface integral over the sphere $S$, 
\beq
\QE
\,:=\,
\frac1{4\pi}\oint_S \Fabu dS_\albe, 
\label{eq:QE}
\eeq
and the condition (\ref{eq:epotconst}) is rewritten 
for $-k^\alpha A_\alpha =\PhiE + C$ with
\beq
C \,=\, -\,\frac1{4\pi\,\QE}\oint_S \PhiE\Fabu dS_\albe.
\eeq

\subsection{First law for systems with a single Killing vector}
\label{sec:1stlaw}

\subsubsection{Definition of the Noether charge $Q$}

Given a 1-parameter family of magnetized perfect-fluid Einstein-Maxwell 
spacetimes specified by 
\beq 
{\cal Q}(\lambda) := [g_{\alpha\beta}(\lambda), u^\alpha(\lambda),  
\rho(\lambda), s(\lambda), A_\alpha(\lambda), j^\alpha(\lambda)],   
\label{eq:variables}
\eeq 
a family of Noether charges is defined on any sphere $S$ that encloses all black holes 
and neutron stars associated with the electric charge and current
\cite{ss77,wald,iyerwald,iyer,sorkin91,brown} :
\beq 
Q = \oint_S Q^{\alpha\beta}\dSab ,
\label{eq:Q}
\eeq 
where 
\beq 
Q^{\alpha\beta} 
= - \frac{1}{8\pi}\na^\alpha k^\beta 
 + k^\alpha \hB^\beta - k^\beta \hB^\alpha,  
\eeq 
and $\hB^\alpha(\lambda)$ is any family of vector fields that satisfies
\beq 
\frac1{\sqrt{-g}}\frac d{d\lambda}(\hB^\alpha\sqrt{-g})  = \Theta^\alpha,   
\eeq 
$\Theta^\alpha$ being defined by Eq.~(\ref{eq:def_Theta}) 
in Appendix~\ref{s:Variation_Lagrangian}.
The vector $\hB^\alpha(\lambda)$ is written, 
\beqn
\hB^\alpha (\lambda)
&=& \frac1{16\pi}(g^{\alpha\gamma}g^{\beta\delta} 
                 -g^{\alpha\beta}g^{\gamma\delta})|_{\lambda=0}  
                 \zna_\beta g_{\gamma\delta}(\lambda) 
\nonumber\\
&&\!\!\!\!\!\!\!\!\!\!\!\!\!\!\!\!
+\,\frac1{4\pi}F^{\beta\alpha}|_{\lambda=0} 
\Big[\,A_\beta(\lambda) - b A_\beta(0)\,\Big]
\,+\,\rm O(\lambda^2),  \ 
\label{eq:b}
\eeqn 
where $\zna_\beta$ is the
covariant derivative of the metric $g_{\alpha\beta}(0)$
and $b$ is a fixed parameter.  

We choose $\hB^\alpha(\lambda)$ to make 
$Q(\lambda)$ finite; and, as we will see below, 
$Q(\lambda)$ is independent of 
the sphere $S$, as long as $S$ encloses the fluid and 
black holes associated with electric charge and current. 
We first choose the parameter $b$ in 
definition (\ref{eq:b}) to have $Q(0)$ satisfy these 
properties.  Regardless of the choice of $\hB^\alpha(0)$, 
the variation of the Noether charge $\dl Q$ is 
finite and independent of the sphere $S$.  
The change in the Noether charge $\dl Q$ results in 
the first law for the Einstein-Maxwell spacetimes with 
charged and magnetized perfect fluid and 
Killing horizons, associated with a single Killing vector 
to impose the stationarity of the system.  

In the calculation of the variation $\dl Q$, 
the Eulerian change of each quantity in Eq.~(\ref{eq:variables}) 
is defined by $\delta {\cal Q} := \frac d{d\lambda} {\cal Q}(\lambda)$, 
and the Lagrangian change at $\lambda=0$ is given by 
\beq 
\Delta  {\cal Q} = (\delta + \Lie_\xi ){\cal Q}, 
\eeq  
where 
$\xi^\alpha$ is a Lagrangian displacement.  
The definition of Lagrangian perturbations is given in 
Appendix \ref{secApp:Lag}.

\subsubsection{Independence of $Q(0)$ on the location of $S$}
\label{sec:Q0}

When the electromagnetic field satisfies $\Fabd \Fabu=0$ 
in the region where the sphere $S$ is located, 
$b=1$ is chosen in Eq.~(\ref{eq:b}) to make 
$Q(0)$ finite and independent of $S$.  In this case, 
we have $\hB^\alpha(0) = 0$. 
When the steady electromagnetic radiation is propagating 
everywhere in the spacetime, $b=1/2$
is chosen.  Then, $\hB^\alpha (0)$ becomes 
$\dis \hB^\alpha (0) 
= F^{\beta\alpha}|_{\lambda=0} A_\beta(0) /8\pi$.  
For the former case, a contribution from the gravitational 
radiation field to the charge $Q(0)$ is subtracted, and 
for the latter case, contributions from the gravitational 
{\it and} electromagnetic radiation fields to the charge $Q(0)$ 
are subtracted; $Q(0)$ is finite and independent of $S$ 
as long as it contains the fluid and all black holes in both cases.

To prove that the charge $Q=Q(0)$ is independent of the sphere $S$, 
we write $Q=Q_K+Q_L$, where $Q_K$ is the Komar charge and $Q_L$ 
an additional contribution related to the surface term 
of the Lagrangian, 
with
\beqn
Q_K &=& -\frac{1}{8\pi} \oint_S \nabla^\alpha
k^\beta dS_{\alpha\beta},
\label{eq:QK}
\\
Q_L 
&=& \oint_S (k^\alpha \hB^\beta - k^\beta \hB^\alpha
)dS_{\alpha\beta} ,
\label{eq:QL}
\eeqn
and rewrite $Q$ in terms of integrals over a spacelike 
hypersurface $\Sigma$ transverse to $k^\alpha$.  
The boundary of $\Sigma$, 
\beq
\pa\Sigma = S\cup_i {\cal B}_i, 
\eeq
is the union of the sphere $S$ and black hole boundaries 
${\cal B}_i$, which is the $i$th connected component 
of $\Sigma \cap \Horiplus$.  
Correspondingly, surface integrals over the $i$th black 
hole horizon ${\cal B}_i$ are written, 
\beqn
Q_{Ki} &=& -\frac{1}{8\pi} \oint_{{\cal B}i} \nabla^\alpha
k^\beta dS_{\alpha\beta},
\\
Q_{Li} 
&=& \oint_{{\cal B}i} (k^\alpha \hB^\beta - k^\beta \hB^\alpha
)dS_{\alpha\beta} .
\eeqn
Then, from the identity 
\beq
\nabla_\beta\nabla^\alpha k^\beta = \Rab k^\beta,
\label{eq:kill}
\eeq 
we have
\beqn
&&
Q_K -\sum_i Q_{Ki} 
\nonumber\\
&=& -\frac{1}{8\pi} \int_{\pa\Sigma} \nabla^\alpha
k^\beta dS_{\alpha\beta} 
\,=\, -\frac{1}{8\pi} \int_\Sigma \Rab k^\beta dS_\alpha 
\nonumber\\
&=& - \frac{1}{8\pi} \int_\Sigma \Gab k^\beta dS_\alpha 
        - \frac{1}{16\pi} \int_\Sigma R k^\alpha dS_\alpha, 
\label{eq:komar}
\eeqn
where the integral over the boundary $\pa\Sigma$ is related 
to the surface integrals with the orientations, 
$
\int_{\pa\Sigma}Q^\albe \dSab
\,=\, \big(\oint_{S}\,-\,\sum\limits_i\oint_{{\cal B}i}\big)Q^\albe \dSab .  
$
If $\Fabd\Fabu=0$ is satisfied in the neighborhood and outside of 
the sphere $S$, the vacuum Einstein equation is satisfied in 
the same region. From 
Eq.~(\ref{eq:komar}) and the choice $\hB^\alpha(0)=0$, 
$Q$ is then independent of the location of $S$.  
For the case $\Fabd\Fabu\ne 0$, using 
\beqn
&& Q_L - \sum_i Q_{Li} 
\nonumber \\
&&
\,=\,\int_{\Sigma} 
\na_\beta(k^\alpha \hB^\beta-k^\beta \hB^\alpha) \, dS_\alpha
\,=\, \int_{\Sigma} 
\na_\beta \hB^\beta k^\alpha dS_\alpha
\nonumber\\
&&
\,=\, \int_{\Sigma} 
\left(\frac1{8\pi}\na_\beta F^\albe A_\alpha
-\frac1{16\pi}\Fabu \Fabd \right) k^\gamma dS_\gamma, 
\label{eq:QLinQ}
\eeqn
we have 
\beqn
&&
Q\,-\,\sum_i Q_i 
\nonumber\\
&=& - \frac{1}{8\pi} \int_\Sigma (\Gab - 8\pi \TabF) k^\beta dS_\alpha 
        - \frac{1}{16\pi} \int_\Sigma R k^\alpha dS_\alpha. 
\nonumber\\
&&
\,+\, \int_{\Sigma} 
\left(\frac1{8\pi}\na_\gamma F^{\beta\gamma} A_\beta k^\alpha
\,-\,\frac1{4\pi} k^\gamma A_\gamma \na_\beta \Fabu  \right) dS_\alpha
\nonumber\\
&&
-\,\sum_i
\frac1{4\pi} \oint_{{\cal B}i}
k^\gamma A_\gamma F^{\alpha\beta} dS_\albe .  
\label{eq:calcQ}
\eeqn
where 
$\TabF$, the stress-energy tensor of 
the electromagnetic field, is defined by Eq.~(\ref{eqApp:TabFu}).  
To derive Eq.~(\ref{eq:calcQ}), we have used the Cartan identity (\ref{eq:Cartan}), 
the symmetry relation $\Lie_k A_\alpha = 0$, 
and Eq.~(\ref{eq:epotconst}).  
From Eq.~(\ref{eq:calcQ}), it is obvious that $Q$ does not 
depend on the sphere $S$ as long as it encloses
all black holes and neutron stars; all integrands of 
the volume integrals over $\Sigma$ in Eq.~(\ref{eq:calcQ}) 
are zero in the region where there are no matters 
and currents, where the sphere $S$ is placed.
This argument may be clearer 
by using an expression of the Komar charge 
associated with the Lagrangian, Eq.~(\ref{eq:QK_Aj}), 
given in Appendix \ref{secApp:dlQ}.

\subsubsection{First law for the charge $Q$}
\label{sec:delQ}

The generalized first law will be obtained by evaluating 
the variation $\dl Q$ in the Noether charge 
in terms of perturbations of the baryon mass, entropy, 
circulation and electric current of each fluid element, 
and the surface areas and charges of the black holes.  
To find the change $\delta Q$, 
we first compute the difference, 
\beq
\delta \Big(Q - \sum_i Q_i \Big), 
\label{eq:dlQ_vol}
\eeq 
between the charge on the sphere $S$ and the sum of the charges 
on the black holes ${\cal B}_i$.  The calculation is performed 
in Appendix \ref{secApp:dlQ} and results in Eq.~(\ref{eqApp:dlQ}).  
In computing the difference (\ref{eq:dlQ_vol}), we choose 
two kinds of gauge: the first one is to choose $\dl k^\alpha = 0$ 
using the diffeomorphism gauge freedom, and the second one 
$\xi^t = 0$ for the Lagrangian displacement as a result of 
the trivial displacement 
(see Appendix~\ref{secApp:dlQ} and FUS).  
For a perfect fluid spacetime, it has been shown in FUS that 
the quantity (\ref{eq:dlQ_vol}) is invariant under gauge 
transformations that respect the Killing symmetry.  
For the case with an electromagnetic field, the same 
invariance under gauge transformations associated with 
diffeomorphisms and the $U(1)$ gauge symmetry 
is shown to hold for the charge $Q$ with a contribution from 
the electromagnetic fields, as is discussed below.

In the black-hole charges $Q_i = Q_{Ki} + Q_{Li}$, 
$Q_{Ki}$ is calculated in FUS: 
\beq
Q_{Ki}\,=\,-\frac1{8\pi}\oint_{{\cal B}i}\na^\alpha k^\beta dS_\albe
 \,=\,\frac1{8\pi} \kappa_i {\cal A}_i, 
\label{eq:Q_Ki}
\eeq
where ${\cal A}_i$ is the area of the $i$th black hole.  
The $Q_{Li}$ is made of contributions from the geometry, 
electric charge, and electromagnetic field.  The former 
has been evaluated in FUS following \cite{bch73}:
\beqn
\dl Q_{Li}
&=&
\oint_{{\cal B}i}(k^\alpha \Theta^\beta - k^\beta \Theta^\alpha)dS_\albe
\nonumber\\
&=&
-\frac1{8\pi}\dl\kappa_i {\cal A}_i 
\,+\,\frac1{4\pi}\oint_{{\cal B}i} k_\alpha F^{\beta\alpha}\dl A_\beta d{\cal A}.
\eeqn
For the latter contribution, since 
$k^\alpha F_{\beta\alpha} = E_\beta$ is parallel to the null generator 
$k_\beta$ on $\Horiplus$, we have
\beqn
k_\alpha F^{\beta\alpha}\dl A_\beta  d{\cal A}
&=& 
k^\alpha F_{\beta\alpha}g^{\beta\gamma}\dl A_\gamma  d{\cal A}
\nonumber
\\
&=& 
k^\alpha F_{\beta\alpha}(-k^\beta \hn^\gamma-\hn^\beta k^\gamma)
\dl A_\gamma  d{\cal A}
\nonumber
\\
&=& 
k^\alpha \hn^\beta \Fabd\dl (k^\gamma A_\gamma) d{\cal A}
\nonumber
\\
&=& 
\dl (k^\gamma A_\gamma)
\Fabu\frac12(k_\alpha \hn_\beta-k_\beta \hn_\alpha) d{\cal A}  
\nonumber
\\
&=& 
\dl (k^\gamma A_\gamma)\Fabu dS_\albe
\eeqn
where $\hn^\alpha$ is the unique null vector field orthogonal 
to ${\cal B}_i$ satisfying $\hn_\alpha k^\alpha = -1$, and 
$\dl k^\gamma = 0$ is used.  
Hence
\beq
\dl Q_{Li}
\,=\,
-\frac1{8\pi}\dl\kappa_i {\cal A}_i 
\,+\,\frac1{4\pi}\oint_{{\cal B}i}
\dl (k^\gamma A_\gamma)\Fabu dS_\albe
\label{eq:dlQ_Li}
\eeq

The contributions from the horizon are Eqs.
 (\ref{eq:Q_Ki}) and (\ref{eq:dlQ_Li}), and 
the surface integral in the r.h.s. of Eq.~(\ref{eqApp:dlQ}), 
\beq
\,-\,
\sum_i \frac1{4\pi}\, \dl\oint_{{\cal B}i}
k^\gamma A_\gamma F^{\alpha\beta} dS_{\alpha\beta}. 
\label{eq:surfQE}
\eeq
Hence the sum of Eqs.~(\ref{eq:dlQ_Li}), (\ref{eq:surfQE}) and 
the perturbed (\ref{eq:Q_Ki}) becomes 
\beqn
&&
\dl Q_i 
\,-\,
\frac1{4\pi}\, \dl\oint_{{\cal B}i}
k^\gamma A_\gamma F^{\alpha\beta} dS_{\alpha\beta}, 
\nonumber
\\
&=&
\frac1{8\pi}\kappa_i \dl{\cal A}_i 
\,-\,\frac1{4\pi}\oint_{{\cal B}i}
k^\gamma A_\gamma \dl(\Fabu dS_\albe)
\nonumber
\\
&=&
\frac1{8\pi}\kappa_i \dl{\cal A}_i 
\,+\,\PhiEi \,\dl \QEi, 
\label{eq:dlQi}
\eeqn
where the total electric charge of the system 
Eq.~(\ref{eq:QE}) is rewritten using Stokes' theorem:  
\beq
\QE
\,=\,
\int_\Sigma j^\alpha \dSa
\,+\,
\sum_i \frac1{4\pi}\oint_{{\cal B}i} \Fabu dS_\albe, 
\label{eq:QEvol}
\eeq
and the electric charge on each black hole is defined by 
\beq
\QEi := \frac1{4\pi}\oint_{{\cal B}i} \Fabu dS_\albe. 
\label{eq:QEi}
\eeq
Note that $\PhiEi$ is defined on each ${\cal B}_i$ by 
\beq
\PhiEi = -A^\alpha k_\alpha = \PhiE + C
\eeq
and is constant.  

Finally, 
when Einstein's equation, Maxwell's equations, their 
linear perturbations and the equation of motion are 
all satisfied, the first law, which relates the change of 
the Noether charge to changes in the thermodynamic and hydrodynamic 
equilibrium of matter, in the electric current and electromagnetic 
field, and in the area and electric charge of the horizon, 
is derived from Eqs.~(\ref{eqApp:dlQ}) and (\ref{eq:dlQi}), 
\beqn
\delta Q  
&=& 
\int_\Sigma \left\{\,\frac{T}{u^t} \Dl (s \, \rho u^\alpha\, \dSa)
\right.
\,+\,\frac{h-Ts}{u^t}\Dl(\rho u^\alpha\, \dSa)
\nonumber \\ 
&&\quad
\,+\,v^\beta \Dl(hu_\beta\,\rho u^\alpha\,\dSa) 
\,-\, A_\beta k^\beta \, \Dl (j^\alpha \dSa)
\nonumber \\ 
&&\left.\phantom{\frac12}
\,-\,(j^\alpha k^\beta - j^\beta k^\alpha )
\Dl A_\beta \,\dSa
\,\right\}
\nonumber \\ 
&+&\sum_i \left(\frac1{8\pi}\kappa_i \dl {\cal A}_i
\,+\, \PhiEi \,\dl \QEi \right).
\label{eq:1stlaw_org}
\eeqn
Here $T$ is the temperature, $s$ the entropy per baryon, 
$h$ the relativistic enthalpy and $v^\alpha$ is defined by the following decomposition 
of the fluid 4-velocity with respect to the helical vector: 
\beq
u^\alpha = u^t(k^\alpha + v^\alpha) \quad\mbox{with}\quad v^\alpha\na_\alpha t = 0 .
\label{eq:4v_decomp}
\eeq 
Note that $u^t=u^\alpha \na_\alpha t$.  
In the special case of stationary and axisymmetric spacetimes (for which $k^\alpha$ is a linear 
combination of the stationary Killing vector and the axisymmetric one), 
Eq.~(\ref{eq:1stlaw_org}) reduces to the mass 
variation formula derived by Carter 
\cite{Carter73,Carter79}. 
\footnote{To derive the mass variation formula for stationary 
and axisymmetric spacetimes from Eq.~(\ref{eq:1stlaw_org}), 
one can replace the helical vector $k^\alpha$ by the timelike 
killing vector $t^\alpha$.  All calculations above are valid 
with this replacement, and now $\dl Q$ becomes $\dl \Madm$
as the sphere $S$ goes to infinity 
(see Sec.~\ref{sec:asymptotic} and FUS).  
Extra terms relating to the 
angular momentum of the fluid and black hole (geometry and 
electromagnetic field) appear in the r.h.s of Eq.~(\ref{eq:1stlaw_org})
as a result of this replacement.}

As mentioned earlier, we can verify now that $Q(\lambda)$ is 
independent of the location of the 2-surface $S$ on which it is 
evaluated.  In Sec.~\ref{sec:Q0}, the charge $Q(\lambda)$ at 
$\lambda=0$ is shown to be independent of $S$, and the variation 
formula Eq.~(\ref{eq:1stlaw_org}) imply that $dQ/d\lambda = \dl Q$ 
is independent of $S$ as long as it encloses the fluid 
and black holes.  

\subsubsection{Gauge invariance of $\dl Q$}

As shown in FUS, for perfect-fluid spacetimes, the difference 
$\dl (Q-\sum\limits_i Q_i)$ is invariant under any gauge transformation 
associated with diffeomorphisms 
that respects the symmetry $k^\alpha$; 
in fact $\dl (Q_K-\sum\limits_i Q_{Ki})$ and $\dl (Q_L-\sum\limits_i Q_{Li})$ are 
separately invariant \cite{Friedman:2001pf}.  
Because of the contribution from the electric potential at the 
horizon, $\delta (Q-\sum\limits_i Q_i)$ is no longer 
invariant when an electromagnetic field is present, 
neither is each contribution.  
We find, however, that an expression in which the contribution of the electric charge 
times the potential at the boundary is subtracted, 
\beq
\dl \Big(Q \,-\, \sum_i Q_i 
\,-\,\frac1{4\pi}\int_{\pa \Sigma}
k^\gamma A_\gamma F^{\alpha\beta} dS_{\alpha\beta} \Big), 
\label{eq:dlQ_inv}
\eeq
is invariant under the gauge transformation that respects 
the symmetry and the $U(1)$ gauge transformation as shown below.

The gauge transformation associated with a vector field $\eta^\alpha$ 
is given by 
\beq
\dl_\eta {\cal Q} \,=\, \Lie_\eta {\cal Q}, \quad 
\xi^\alpha(\eta) \,=\, -\eta^\alpha, 
\eeq
and the corresponding Lagrangian variation is identically zero, 
\beq
\Dl_\eta \,=\, \dl_\eta + \Lie_{-\eta} \,=\, 0.
\eeq
We decompose the vector $\eta^\alpha$ with respect to the symmetry 
$k^\alpha$, 
\beq
\eta^\alpha \,=\, \eta^\alpha \na_\alpha t \,\, k^\alpha 
\,+\, \hat{\eta}^\alpha ,
\eeq
with $\hat{\eta}^\alpha \na_\alpha t = 0$.  

Then, the change in $\dl (Q_L \,-\, \sum\limits_i Q_{Li})$ 
becomes 
\beqn 
&&
\dl_\eta \Big( Q_L -\sum_i Q_{Li}\Big)
\,=\, 
\int_\Sigma\na_\beta\Theta^\beta k^\alpha\dSa 
\,=\, 
\int_\Sigma \dl_\eta \Lag \,d^3x
\nonumber\\
&&
\,=\, 
\int_\Sigma \na_\alpha (\Lag \,\hat{\eta}^\alpha) d^3x
\,=\, 
-\frac1{8\pi}\int_{\pa\Sigma} F_{\gamma\delta} F^{\gamma\delta}\,
k^\alpha\hat{\eta}^\beta\,\dSab, 
\nonumber\\
\label{eq:QLgauge}
\eeqn 
where we used the relation 
$\dl_\eta \Lag = \Lie_\eta \Lag = 
\na_\alpha (\Lag \,\hat{\eta}^\alpha)= 
\na_\beta (\Lag \,k^\alpha\na_\alpha t\,\hat{\eta}^\beta)$, 
with $k^\alpha\na_\alpha t =1$. The non-zero contribution 
to the Lagrangian density $\Lag$ at the boundary $\pa\Sigma$ 
is that of the electromagnetic field $\LagF$.  

Similarly $\dl (Q_K \,-\, \sum\limits_i Q_{Ki} )$ is calculated 
from Eq.~(\ref{eq:komar}):
\beqn
&&
\dl_\eta \Big( Q_K -\sum_i Q_{Ki}\Big)
\,=\, 
-\,\frac1{8\pi} \dl_\eta \int_\Sigma\Rab \,k^\beta\dSa 
\nonumber\\
&&
\,=\, 
-\,\frac1{8\pi} \int_{\pa\Sigma} 2R^\alpha{}_\gamma \,
k^\gamma\hat{\eta}^\beta\,\dSab
\nonumber\\
&&
\,=\, 
-\,\frac1{2\pi} \int_{\pa\Sigma}
\left[\,\Lie_k A_\delta - \na_\delta(k^\gamma A_\gamma)\,\right]
F^{\alpha\delta}\hat{\eta}^\beta\,\dSab
\nonumber\\
&& \quad\ 
+\,\frac1{8\pi}\int_{\pa\Sigma} F_{\gamma\delta} F^{\gamma\delta}\,
k^\alpha\hat{\eta}^\beta\,\dSab, 
\label{eq:QKgauge}
\eeqn
where we have substituted $\Rab = 8\pi\TabF$ at $\pa\Sigma$
and Eq.~(\ref{eqApp:TabFu}), before using Eq.~(\ref{eq:Cartan}).  
Finally, the last term in Eq.~(\ref{eq:dlQ_inv}) becomes
\beqn
&&\quad \ 
\,-\,\frac1{4\pi}\dl_\eta \int_{\pa \Sigma}
k^\gamma A_\gamma F^{\alpha\beta} dS_{\alpha\beta} 
\nonumber\\
&&
\,=\,
\,-\,\frac1{4\pi}\dl_\eta \int_{\Sigma}
\na_\delta(k^\gamma A_\gamma F^{\alpha\delta}) \dSa
\nonumber\\
&&
\,=\,
\,-\,\frac1{2\pi}\int_{\pa \Sigma}
\na_\delta(k^\gamma A_\gamma F^{\alpha\delta})\hat{\eta}^\beta \dSab.  
\label{eq:Qelegauge}
\eeqn
Adding Eqs.~(\ref{eq:QKgauge}), (\ref{eq:QLgauge}), 
and (\ref{eq:Qelegauge}), and imposing $\Lie_k A_\alpha = 0$ 
and $\na_\beta \Fabu = 0$ at $\pa \Sigma$, all terms cancel out :
\beq
\dl_\eta \Big(Q \,-\, \sum_i Q_i 
\,-\,\frac1{4\pi}\int_{\pa \Sigma}
k^\gamma A_\gamma F^{\alpha\beta} dS_{\alpha\beta} \Big)\,=\,0. 
\eeq
Hence the difference (\ref{eq:dlQ_inv}) 
is invariant under a gauge transformation that respects 
the symmetry.

For the $U(1)$ gauge transformation, 
we directly show, instead of Eq.~(\ref{eq:dlQ_inv}), 
the invariance of the difference evaluated at the surface $S$, 
\beq
\dl \Big(Q 
\,-\,\frac1{4\pi}\oint_{S}
k^\gamma A_\gamma F^{\alpha\beta} dS_{\alpha\beta} \Big), 
\label{eq:dlQ_inv_surf}
\eeq
under the transformation with a gauge potential $f$, 
\beq
\dl_f A_\alpha = \na_\alpha f.
\eeq
The change in charge $Q$ with this transformation is 
\beq
\dl_f Q \,=\, \dl_f \QL 
\,=\, \frac1{4\pi}\oint_S 
(k^\alpha F^{\gamma\beta} - k^\beta F^{\gamma\alpha})
\dl_f A_\gamma dS_{\alpha\beta}.
\eeq
Then, the difference (\ref{eq:dlQ_inv_surf}) vanishes
\beqn
&&
\dl_f \Big(Q 
\,-\,\frac1{4\pi}\int_{S}
k^\gamma A_\gamma F^{\alpha\beta} dS_{\alpha\beta} \Big) 
\nonumber\\
&&
\,=\,
-\frac3{4\pi}\oint_S k^{[\alpha}F^{\beta\gamma]}\na_\gamma f \,
 dS_{\alpha\beta}
 \,=\,0, 
\eeqn
because integration by part of the r.h.s.~of the first equality
becomes an itegration of a divergence over $S$ that vanishes, 
and an integral of 
\beqn
3\na_\gamma (k^{[\alpha}F^{\beta\gamma]})
&=& 
2 k^{[\alpha} \na_\gamma F^{\beta]\gamma}
\,+\,
\na_\gamma k^\gamma F^{\alpha\beta}
\,+\,
\Lie_k F^{\alpha\beta}
\nonumber\\
&=& 0, 
\eeqn
when the Maxwell's equation is satisfied on $S$ where the current 
is zero, and $k^\alpha$ the Killing vector.

\subsubsection{Asymptotically flat systems}
\label{sec:asymptotic}

FUS have derived the first law 
in a Hamiltonian framework, and shown the 
relations between $Q_K$ and $\dl Q_L$
and the asymptotic quantities, 
the ADM mass $\Madm$, the Komar mass $\MK$ associated with 
the timelike asymptotic Killing vector $t^\alpha$, and 
the angular momentum $J$ associated with the asymptotic rotational 
Killing vector $\phi^\alpha$. In the presence of an electromagnetic field, 
the only difference with FUS is the following term in $\dl \QL$ 
\beq
\oint_\infty
 (k^\alpha\ThFbu -k^\beta\ThFau) dS_{\alpha\beta}, 
\eeq
where 
\beq
\oint_\infty \,:=\, \lim_{r\rightarrow\infty}\oint_{S_r}, 
\eeq
with $S_r$ is a sphere of a radius $r$, and 
$\ThFau$ is the surface term of the variation of 
electromagnetic Lagrangian defined by 
\beq
\ThFau\,=\,\frac1{4\pi}F^{\beta\alpha}\dl A_\beta.
\eeq
However, this does not contribute to $\dl Q_L$, because, 
for asymptotically flat systems, the components of 
$A_\alpha$ are $\Od(r^{-1})$ or lower, 
and, accordingly, those of $\Fabu$ are 
$\Od(r^{-2})$ or lower, hence
\beq
\oint_\infty
 (k^\alpha\ThFbu -k^\beta\ThFau) dS_{\alpha\beta}
 \,=\,
\lim_{r\rightarrow\infty}\oint_{S_r}
\ThFau  \na_\alpha r \, r^2 d\Omega
\,=\,0
\eeq
where the relations $k^\alpha \na_\alpha t = 1$ and 
$k^\alpha \na_\alpha r = 0$ have been used.
Therefore, as in FUS, 
\beqn
Q_K 
&=&
-\frac{1}{8\pi} \oint_\infty \nabla^\alpha  k^\beta dS_{\alpha\beta}
\,=\,
\frac12\, M_K -\Omega J
\\
\dl Q_L 
&=&
\oint_\infty (k^\alpha\Theta^\beta -k^\beta\Theta^\alpha) dS_{\alpha\beta}
\nonumber\\
&=&
\dl\Madm \,-\, \frac12 \dl M_K \,+\, \dl \Omega\, J
\eeqn
which results in 
\beq
\dl Q \,=\, \dl \Madm \,-\, \Omega\, \dl J .
\eeq

As we will see below, when two nearby equilibria are compared 
conserving the integral quantities, including the generalized Kelvin circulation
for magnetized flow, and the areas and electric 
charges of the black holes, so that the r.h.s.~of 
Eq.~(\ref{eq:1stlaw_org}) vanishes, 
the first law is simply written $\dl Q = 0$, or 
$ \dl \Madm \,=\, \Omega\, \dl J $
for asymptotically flat systems.

\section{Comparing stationary systems}
\label{sec:conserv}

\subsection{Ideal MHD flow}

\subsubsection{Conservation of rest mass, entropy and electric charge}

For an isentropic fluid, conservation of rest mass and entropy 
can be expressed by means of a Lie derivative along the fluid 4-velocity
$u^\alpha$ : 
\beq
\Lie_u(\rho \sqrt{-g}) \,=\,0, \quad \Lie_u s \,=\, 0
\label{eq:consrhos}
\eeq
and if these quantities are conserved in the perturbed states, 
the perturbed conservation laws have first integrals
\beq
\Dl (\rho\sqrt{-g})\,=\,0, \quad \Dl s\,=\,0,
\label{eq:pconsrhos}
\eeq
Since we assume that the electric current is not necessarily 
colinear to the fluid 4-velocity, conservation of 
 electric current, 
\beq
\Lie_j \sqrt{-g}\,=\,0,
\eeq
does not imply another perturbed conservation law analogous to 
Eq.~(\ref{eq:pconsrhos}) with respect to the lagrange perturbation 
of the congruence of flow lines, that is, $\Dl (j^\alpha \dSa)\neq 0$.  
However, its volume integral 
over the neutron star should vanish because of the conservation 
of electric charge:
\beq
\dl \QEm \,=\,
\dl\int_\Sigma j^\alpha \dSa\,=\,
\int_\Sigma \Dl(j^\alpha \dSa)\,=\,0. 
\eeq

\subsubsection{Conservation of magnetic flux for ideal MHD}
\label{sec:Magflux}

Assuming perfect conductivity for the magnetized flow 
of the neutron star matter, 
the ideal MHD condition
\beq
\Fabd u^\beta \,=\, 0, 
\label{eq:idealMHD}
\eeq
is satisfied, and 
the curl of Eq.~(\ref{eq:idealMHD}) becomes 
\beq
\Lie_u \Fabd = 0 
\label{eq:LieuF}
\eeq
as a result of the Cartan identity and $(dF)_{\albe\gamma}=0$.  
Eq.~(\ref{eq:LieuF}) implies the well known 
conservation law of magnetic flux, Alfven's theorem.  
Let us introduce the map $\psi_\tau$
as the family of diffeomorphisms generated by $u^\alpha$, namely 
the curve $\tau \rightarrow \psi_\tau(P)$ has the
tangent vector $u^\alpha(P)$ at a point $P$.  
For any closed curve $c$ contractable to a point, 
we consider the closed curve 
$c_\tau = \psi_\tau \circ c$ obtained by moving each point of 
$c$ during the proper time $\tau$
along the fluid trajectory through that point. 
Then the conservation of magnetic flux, which is 
the integral form of the law (\ref{eq:LieuF}),
amounts to the conservation of the integral of the 1-form $A_\alpha$ along 
the closed curve $c_\tau$ in the fluid : 
\beq \label{eq:Alfven_theorem}
\oint_{c_\tau} A_\alpha \,d\ell^\alpha \,=\, \CF \,=\, \mbox{const}.
\eeq

When the perturbed state also satisfies the ideal MHD condition, 
the  perturbed version of the conservation of magnetic flux \eqref{eq:variables} has a first integral 
\beq
\Dl \Fabd = 0 
\label{eq:pconsmag}
\eeq 
and hence $\Dl\Fabd = (d\Dl A)_\albe = 0$ 
is satisfied for any region in the fluid.  
For $(d\Dl A)_\albe = 0$ to be satisfied, it suffices that 
$\Dl A_\alpha = \na_\alpha\Psi$ for some scalar field $\Psi$.
Conversely, as long as the fluid support (neutron star) 
is star-convex, the Poincar\'e lemma guarantees 
the existence of $\Psi$.
As a result, 
the last term of the volume integral of 
Eq.~(\ref{eq:1stlaw_org}) vanishes:  
\beqn
&&
\int_\Sigma
\,-\,(j^\alpha k^\beta - j^\beta k^\alpha )
\Dl A_\beta \,\dSa
\nonumber\\
&=&
\int_{\pa \Sigma}
\,-\,(j^\alpha k^\beta - j^\beta k^\alpha )
\Psi \,dS_{\alpha \beta}
\nonumber\\
&+&
\int_{\Sigma}
\na_\beta(j^\alpha k^\beta - j^\beta k^\alpha )
\Psi \,\dSa
\,=\,0, 
\label{eq:jkterm}
\eeqn
because there is no electric current on the boundary surface $\pa \Sigma$, 
and a relation,  
$
\na_\beta(j^\alpha k^\beta - j^\beta k^\alpha )
=\Lie_k j^\alpha + j^\alpha \na_\beta k^\beta - k^\alpha \na_\beta j^\beta=0, 
$
is satisfied for the conserved current $j^\alpha$ that respects the symmetry.

\subsubsection{Conservation of circulation for ideal MHD: generalized 
Kelvin's Theorem}
\label{sec:iMHD_Oron}

When two equilibria of some ideal MHD flow are compared 
with the same rest mass, same entropy and same magnetic flux,
the perturbed conservation laws (\ref{eq:pconsrhos}) and 
(\ref{eq:pconsmag}), as well as Eq.~(\ref{eq:jkterm}) are satisfied.
Then the change in the Noether charge (\ref{eq:1stlaw_org}) becomes 
\beqn
\delta Q
&=& 
\int_\Sigma \left[\,
v^\beta \Dl\left(hu_\beta\,\rho u^\alpha \,\dSa\right) 
\,-\,A_\beta k^\beta \Dl(j^\alpha\, \dSa ) 
\,\right]
\nonumber \\ 
&+&\sum_i \left(\frac1{8\pi}\kappa_i \dl {\cal A}_i
\,+\, \PhiEi \,\dl \QEi \right).
\label{eq:1stlaw_idealMHD}
\eeqn

For some perfect fluid {\it without} magnetic field, 
the circulation of the flow is conserved.  
The curl of the relativistic Euler equation 
$u^\beta \hat \omega_\beal=0$ 
is written 
$
\Lie_u \hat\omega_\albe = 0 
$
where $\hat\omega_\albe$ is the relativistic vorticity 
defined by $ \hat\omega_\albe \,=\, (d(h u))_\albe $
and a corresponding integral law, known as Kelvin's theorem, 
is the conservation of circulation, 
the line integral of $hu_\alpha$ along an arbitrary 
closed curve comoving with the fluid.  
As shown in FUS, the contribution from the circulation 
to the change in the Noether charge $\dl Q$ is included in 
the term 
\beq
\int_\Sigma v^\beta \Dl(hu_\beta\,\rho u^\alpha\,\dSa) , 
\label{eq:circulation}
\eeq
which vanishes when the circulation is conserved 
in the perturbed flow, for example, when the irrotational 
flow, or the corotational flow, is maintained.  
This can be shown in the same way as 
eliminating a term (\ref{eq:jkterm}) using 
the conservation of magnetic flux.

The integral in Eq.~(\ref{eq:1stlaw_idealMHD}), however, 
does not in general vanish for magnetized flows, or
even for ideal MHD flows, 
because of the lack of a conservation of circulation law in the magnetized case.  
This can be seen from the relativistic MHD-Euler equation 
which is not the inner product of the fluid 
4-velocity and an exact two form, because of the Lorenz 
force on the right hand side, 
\beq
u^\beta (d(h u))_\beal = \frac1{\rho}\Fabd j^\beta. 
\label{eq:MHD-Euler}
\eeq

Nevertheless, 
Bekenstein and Oron \cite{Bekenstein:2000sf} 
(see also \cite{BB06}) 
have found that, if the 4-current takes the form 
\beq
j^\alpha 
 \,=\, \na_\beta(\rho u^\alpha q^\beta\,-\,\rho u^\beta q^\alpha), 
\label{eq:BOcur}
\eeq
where $q^\alpha$ is an auxiliary vector field,
one can obtain a generalized conserved circulation for magnetized flow.  
This 4-current is derived from the variation of a Lagrangian 
in which the ideal MHD condition is added as an interaction term 
$\rho q^\alpha \Fabd u^\beta$ with the Lagrange multiplier $\rho q^\alpha$. 
The form (\ref{eq:BOcur}) manifestly satisfies the electric charge conservation: 
$\na_\alpha j^\alpha = 0$. Note that, for a given 4-current $j^\alpha$, one has the degree of 
freedom to change $q^\alpha$ according to 
\beq \label{eq:change_q}
  q^\alpha \mapsto q^\alpha + \lambda u^\alpha
\eeq
for a scalar $\lambda$ without affecting the value of $j^\alpha$.

Using $\nabla_\alpha (\rho u^\alpha) = 0$ [cf. Eq.~(\ref{eq:consrhos})],
the 4-current (\ref{eq:BOcur}) can be rewritten 
\beq
j^\alpha
 \,=\, \Lie_q (\rho u^\alpha) 
 \,+\,\rho u^\alpha \na_\beta q^\beta  .
\label{eq:BOcur2}
\eeq
Substituting the form into the Lorenz force, 
we get
\beq
\frac1{\rho}\Fabd j^\beta 
\,=\, \frac1{\rho} \Fabd \Lie_q(\rho u^\beta)
\,=\, (d\eta)_\albe u^\beta, 
\label{eq:BO_LFterm}
\eeq
where 
1-form $\eta_\alpha$ is defined by 
\beq
\eta_\alpha := \Fabd q^\beta,
\label{eq:defeta}
\eeq
and a relation (\ref{eq:lemma}) from Appendix~\ref{s:cal_BO_LFterm}, 
which is implied by the ideal MHD condition (\ref{eq:idealMHD}), is used. 
Note that, thanks to (\ref{eq:idealMHD}), the 1-form $\eta_\alpha$ 
does not depend on the specific 
choice of $q^\alpha$ within the range allowed by (\ref{eq:change_q}).
By means of (\ref{eq:BO_LFterm}), the MHD-Euler 
equation (\ref{eq:MHD-Euler}) 
is simply written, 
\beq
u^\beta (dw)_\beal \,=\, 0, 
\label{eq:MHD-Euler_form1}
\eeq
where $w_\alpha$ is the generalized momentum 1-form defined by 
\beq
w_\alpha:=hu_\alpha \,+\,\eta_\alpha.
\label{eq:defw}
\eeq
From Eq.~(\ref{eq:MHD-Euler_form1}) one can easily deduce a generalised 
conservation of circulation law for ideal MHD flows. Indeed, defining the 
\emph{vorticity $\oabd$ of a magnetized flow} by 
\beq
\oabd \,=\, \na_\alpha w_\beta - \na_\beta w_\alpha
\,=\, (dw)_\albe,
\label{eq:def_vorticity} 
\eeq
the Cartan identity, combined with Eq.~(\ref{eq:MHD-Euler_form1}) and the identity
$d\omega=d^2 w=0$, yields 
\beq
\Lie_u \oabd = 0 . 
\label{eq:Lieuomega}
\eeq
By means of the Stokes theorem, this conservation law can be put in the following 
integral form
[using the same notation as in Eq.~(\ref{eq:Alfven_theorem})]:
\beq
\oint_{c_\tau} (hu_\alpha +\eta_\alpha)\,d\ell^\alpha \,=\, C_{\rm m} \,=\, \mbox{const}.
\label{eq:MagC}
\eeq
This law, which has been first derived by Bekenstein and Oron \cite{Bekenstein:2000sf}, 
constitutes a generalization to ideal MHD of the relativistic Kelvin's circulation theorem
(which corresponds to $\eta_\alpha=0$, see e.g. \cite{Gourg06})

One can repeat the same argument as for the magnetic flux in 
the previous section.  
The perturbation of Eq.~(\ref{eq:Lieuomega}) for the magnetized vorticity has first integral 
\beq
\Dl \oabd = 0,  
\label{eq:pconsmagcir}
\eeq
which implies $\Dl \oabd = (d\Dl(h u+\eta))_\albe = 0$. The
Poincar\'e lemma guarantees that a function $\Psi$ exists 
on the star-convex fluid support such that 
$\Dl(hu_\alpha+\eta_\alpha) = \na_\alpha \Psi$.  

It is also suggested from Eq.~(\ref{eq:MHD-Euler_form1}) that 
an irrotational magnetohydrodynamic flow, $\omega_{\alpha \beta}=0$, 
is described by a velocity potential $\Phi$ that satisfies 
\beq
hu_\alpha+\eta_\alpha = \na_\alpha \Phi,
\label{eq:irrot}
\eeq
and, in this case, $\Psi = \Dl \Phi$.

\subsubsection{First law for the ideal MHD with Bekenstein - 
Oron current}
\label{sec:1stlaw_BO}

For ideal MHD flow with the Bekenstein-Oron current (\ref{eq:BOcur}), 
the first law of the form Eq.~(\ref{eq:1stlaw_idealMHD}) 
is further simplified when comparing two nearby equilibria that conserve 
the circulation of a magnetized flow, (\ref{eq:MagC}).  
Substituting Eq.~(\ref{eq:BOcur}) to 
the second term of the integrand of the volume integral in 
Eq.~(\ref{eq:1stlaw_idealMHD}), 
we have 
\beqn
&&\!\!\!\!\!\!\!\!\!\!
\,-\, A_\beta k^\beta\, \Dl (j^\alpha \dSa)
\nonumber\\
&&
\,=\,
(\Lie_k A_\gamma -k^\beta F_{\beta\gamma})\,
\Dl \left[(\rho u^\alpha q^\gamma -\rho u^\gamma q^\alpha)\dSa\right]
\nonumber\\
&&
\,-\, \na_\gamma\left\{ A_\beta k^\beta\,
\Dl \left[(\rho u^\alpha q^\gamma -\rho u^\gamma q^\alpha)\dSa\right] \right\}
\nonumber\\
&&
\,=\,
v^\beta\Dl(\eta_\beta \rho u^\alpha \dSa)
\,-\,v^\beta q^\gamma\Dl F_{\beta\gamma}\,\rho u^\alpha \dSa
\nonumber\\
&&
\,-\,\frac1{u^t}u^\beta F_{\beta\gamma}
\,\Dl(\rho u^\alpha q^\gamma\,\dSa)
\,-\,k^\beta F_{\beta\gamma}\Dl (\rho u^\gamma q^\alpha\,\dSa)
\nonumber\\
&&
\,-\, \na_\gamma\left\{ A_\beta k^\beta\,
\Dl \left[(\rho u^\alpha q^\gamma -\rho u^\gamma q^\alpha)\dSa\right] \right\}, 
\label{eq:BBDljdS}
\eeqn
where the relation 
$\Dl \na_\beta (f^{\albe} \dSa) = \na_\beta \Dl (f^{\albe} \dSa)$, valid
for any antisymmetric tensor $f^\albe$, 
and the Cartan identity (\ref{eq:Cartan}) 
are used, and 
the symmetry $\Lie_k A_\gamma = 0$ is imposed.

Since the ideal MHD condition (\ref{eq:idealMHD}) is satisfied, 
terms including $\Fabd u^\beta$ are discarded.  Also 
a term involving $\Fabd \Dl u^\beta$ is discarded, 
because $\Dl u^\beta$ is colinear to $u^\beta$ (see Eq.~(\ref{eq:Dlu})).
Substituting Eq.~(\ref{eq:BBDljdS}) to Eq.~(\ref{eq:1stlaw_org}), 
the integral of the last term of Eq.~(\ref{eq:BBDljdS}) 
becomes a surface integral on $\pa \Sigma$ which vanishes.  
Hence, the first law (\ref{eq:1stlaw_org}) 
for the Bekenstein-Oron formulation of ideal MHD 
is written 
\beqn
\delta Q  
&=& 
\int_\Sigma \left\{\,\frac{T}{u^t} \Dl (s \, \rho u^\alpha\, \dSa)
\right.
\,+\,\frac{h-Ts}{u^t}\Dl(\rho u^\alpha\, \dSa)
\nonumber \\ 
&&\quad
\,+\,v^\beta \Dl\left[(hu_\beta+\eta_\beta)\,\rho u^\alpha\,\dSa\right] 
\nonumber \\ 
&&\quad
\,-\,v^\beta q^\gamma\Dl F_{\beta\gamma}\,\rho u^\alpha \dSa
\nonumber \\ 
&&\left.\phantom{\frac12}
\,-\,(j^\alpha k^\beta - j^\beta k^\alpha )
\Dl A_\beta \,\dSa
\,\right\}
\nonumber \\ 
&+&\sum_i \left(\frac1{8\pi}\kappa_i \dl {\cal A}_i
\,+\, \PhiEi \,\dl \QEi \right).
\label{eq:1stlaw_BBcur}
\eeqn
Introducing the following notation
\beqn
&
dM_{\rm B}:=\rho u^\alpha \dSa, 
\quad
dS:= s\,dM_{\rm B}, 
&
\nonumber\\
&
dC_\alpha:= (h u_\alpha +\eta_\alpha) dM_{\rm B},
&
\eeqn
we further rewrite Eq.~(\ref{eq:1stlaw_BBcur}) as
\beqn
\delta Q
&=& 
\int_\Sigma \left\{\,\frac{T}{u^t}  \Dl dS
\right.
\,+\,\frac{h-Ts}{u^t}\Dl dM_{\rm B}
\,+\,v^\alpha \Dl dC_\alpha
\nonumber \\ 
&&\left.\phantom{\frac12}\!\!\!\!\!\!\!\!\!\!
\,-\,v^\beta q^\gamma\Dl F_{\beta\gamma}\,dM_{\rm B}
\,-\,(j^\alpha k^\beta - j^\beta k^\alpha )
\Dl A_\beta \,\dSa
\,\right\}
\nonumber \\ 
&+&\sum_i \left(\frac1{8\pi}\kappa_i \dl {\cal A}_i
\,+\, \PhiEi \,\dl \QEi \right).
\label{eq:1stlaw_BBcur2}
\eeqn

When the circulation of magnetized flow 
is conserved, there exists a potential $\Psi$ such that 
$\Dl (h u_\alpha+\eta_\alpha) = \na_\alpha \Psi$.  
Applying an argument analogous to that for the magnetic 
flux in Sec.~\ref{sec:Magflux}, a term for 
the circulation of magnetized flow in 
the r.h.s. of Eq.~(\ref{eq:1stlaw_BBcur}) vanishes:
\beqn
&&
\int_\Sigma
v^\beta \Dl(hu_\beta+\eta_\beta)\,\rho u^\alpha\,\dSa 
\nonumber\\
&=&
\int_\Sigma
(\rho u^\alpha v^\beta - \rho u^\beta v^\alpha) \na_\beta \Psi \,\dSa 
\,=\,0
\label{eq:circterm}
\eeqn
where $v^\alpha dS_\alpha = 0$ is used in the first equality, 
and the last equality is proved in the same way as in Eq.~(\ref{eq:jkterm}) 
because of a relation, 
$
\na_\beta(\rho u^\alpha v^\beta - \rho u^\beta v^\alpha )
\,=\,
\na_\beta(\rho u^\beta k^\alpha - \rho u^\alpha k^\beta)
\,=\,
\,-\,\Lie_k (\rho u^\alpha) 
\,-\, \rho u^\alpha \na_\beta k^\beta 
\,+\, k^\alpha \na_\beta (\rho u^\beta)
\,=\,0.
$
Therefore, 
the rest mass, entropy, circulation of magnetized flow and magnetic flux 
are all conserved in the perturbation of ideal MHD flow 
with the Bekenstein-Oron current (\ref{eq:BOcur}), 
namely, Eqs.~(\ref{eq:pconsrhos}), (\ref{eq:pconsmag}), and 
(\ref{eq:pconsmagcir}) are satisfied, 
the change in the Noether charge (\ref{eq:1stlaw_BBcur}) becomes 
\beq
\dl Q
\,=\, 
\sum_i \left(\frac1{8\pi}\kappa_i \dl {\cal A}_i
\,+\, \PhiEi \,\dl \QEi \right).
\label{eq:1stlaw_idealBBcur}
\eeq

\subsection{MHD flow without conduction current}

It is expected that the inner core of the neutron star may be
composed of a mixture of superfluid protons and high-energy 
particles.  Such flows are well described by an ideal fluid without 
conduction current but only convection current:
\beq 
j^\alpha = \rho e u^\alpha,
\label{eq:curNC}
\eeq
where $e$ is the electric charge per baryon mass \cite{OronBB}.  
Conservation of rest mass, $\na_\alpha (\rho u^\alpha) = 0$, and 
 current, $\na_\alpha j^\alpha = 0$, imply that the 
specific charge $e$ is conserved along fluid flow lines, 
\beq
\Lie_u e \,=\, 0.
\label{eq:conscharge_NC}
\eeq
Substituting the current (\ref{eq:curNC}) into the first law 
(\ref{eq:1stlaw_org}), we have 
\beqn
&& \!\!\!\!
\delta Q  
\,=\, 
\int_\Sigma \left\{\,\frac{T}{u^t} \Dl (s \, \rho u^\alpha\, \dSa)
\right.
\,+\,\frac{h-Ts}{u^t}\Dl(\rho u^\alpha\, \dSa)
\nonumber \\ 
&& \!\!\!\!
\,+\,\left. v^\beta \Dl[(hu_\beta+eA_\beta)\rho u^\alpha\,\dSa] 
\,-\, \frac{A_\beta u^\beta}{u^t} \, \Dl (e \rho u^\alpha \dSa)
\right\}
\nonumber \\ 
&& \!\!\!\!
\,+\,\sum_i \left(\frac1{8\pi}\kappa_i \dl {\cal A}_i
\,+\, \PhiEi \,\dl \QEi \right), 
\label{eq:1stlaw_NC}
\eeqn
and also into the MHD-Euler equation (\ref{eq:MHD-Euler}), 
\beq
u^\beta (d(hu + eA))_\beal \,+\, A_\beta u^\beta \na_\alpha e \,=\,0.  
\label{eq:MHD-Euler_NC}
\eeq

As shown in \cite{OronBB}, the circulation of 
the magnetized flow defined by 
\beq
\Gamma : = \oint_{c_\tau} (hu_\alpha + e A_\alpha)\,d\ell^\alpha.
\label{eq:MagC_NC}
\eeq
is conserved only when the closed curve $c_\tau$ is taken 
along a curve of constant specific charge $e$.  
If we further assume that the charge is distributed initially 
satisfying 
\beq
e = e(A_\alpha u^\alpha) 
\eeq
(or merely $e= \rm constant$ in the simplest case), 
the curl of Eq.~(\ref{eq:MHD-Euler_NC}) becomes a law of conservation of circulation for magnetized flow, 
\beq
\Lie_u (d(hu + eA))_\beal = 0, 
\eeq
and $\Gamma$ is constant for any closed curved $c_\tau$ comoving with the flow.  
Then, with the same argument in Sec.~\ref{sec:1stlaw_BO}, when nearby 
equilibrium solutions having the same value of circulation 
$\Gamma$ are compared, the perturbed conservation law, 
\beq
\Dl (d(hu + eA))_\beal = 0, 
\label{eq:pconscirc_NC}
\eeq
is satisfied.  
Hence, with Eq.~(\ref{eq:pconscirc_NC}), 
a perturbation of Eq.~(\ref{eq:conscharge_NC}), 
\beq
\Dl e = 0, 
\label{eq:pconscharge_NC}
\eeq
and conservation of rest mass and entropy (\ref{eq:pconsrhos}), 
the first law for a flow without conduction current 
is also written simply as Eq.~(\ref{eq:1stlaw_idealBBcur}).  
It should be noted that the condition $e= \rm constant$ may not 
be too restrictive for an application such as the superfluid 
proton component in a neutron star interior.

\section{Integrability condition for the MHD-Euler 
equation in ideal MHD}
\label{sec:integrability}

When the stationarity or helical symmetry is imposed 
explicitly on the (MHD-)Euler equation, it is 
no longer an evolution equation.  
In usual methods \cite{BNSCF,UryuCFWL,BHNS_QE}, 
its numerical solution is calculated using 
its first integral - a sufficient condition 
for the stationary or helically symmetric 
(MHD-)Euler equation being satisfied.  
Therefore, finding the first integral 
is a key, and also a restriction, 
for computing equilibrium solutions considered in 
Sec.~\ref{sec:conserv}.

As shown in Sec.~\ref{sec:iMHD_Oron}, when the Bekenstein-Oron 4-current 
(\ref{eq:BOcur}) is introduced, the relativistic MHD-Euler equation 
for ideal MHD flows takes the form (\ref{eq:MHD-Euler_form1}).  
If we assume that the generalized momentum (\ref{eq:defw}) of 
the magnetized flow respects the helical symmetry, 
$\Lie_k w_\alpha = 0$, then 
a first integral is immediately 
derived for corotational and irrotational flows, in a way 
fully analogous with the non-magnetized case 
\cite{irbns_formulation} (see also \cite{Gourg06}):
the Cartan identity $k^\beta \omega_{\beta\alpha} 
= \Lie_k w_\alpha-\na_\alpha(w_\beta k^\beta) $
reduces to $k^\beta \omega_{\beta\alpha} = - \na_\alpha(w_\beta k^\beta)$
and, for an irrotational flow ($\omega_{\beta\alpha} =0$), 
or for a corotational one [$u^\alpha$ colinear to $k^\alpha$ so that
(\ref{eq:MHD-Euler_form1}) implies
$k^\beta \omega_{\beta\alpha} =0$], we get the first integral 
$w_\alpha k^\alpha = \mathrm{const}$.

However, it turns out that the assumption $\Lie_k w_\alpha = 0$
is too restrictive when applied to a corotating flow.  
In view of (\ref{eq:defw}) and (\ref{eq:defeta}), it would yield the 
first integral
$ w_\alpha k^\alpha = 
h u_\alpha k^\alpha + F_{\alpha\beta} k^\alpha q^\beta = \mathrm{const}$.
Now, the colinearity of $k^\alpha$ and $u^\alpha$, along with
the ideal MHD condition (\ref{eq:idealMHD}), implies 
$\Fabd k^\beta = 0$. Hence the first integral would reduce to 
$h u_\alpha k^\alpha = \mathrm{const}$, i.e. exactly the same as in 
the perfect fluid case, without any Lorentz force term.

In Bekenstein and Oron's theory \cite{Bekenstein:2000sf,BB06}, 
the momentum $w_\alpha$ defined by (\ref{eq:defw}) and 
(\ref{eq:defeta}) contains the Lagrange multiplier $q^\alpha$.  
Because $q^\alpha$ is not a physical quantity, it does not necessarily 
obey the helical symmetry.  
This has been noticed by Bekenstein and Oron, but has not 
been taken into account when the first integral was derived.
In this section, we briefly review the properties of 
the 4-current by Bekenstein and Oron, then 
derive integrability conditions 
for the case when $q^\alpha$ does not respect the symmetry.

A first integral for an axisymmetric and rigidly rotating 
neutron star has been derived by Bonazzola, Gourgoulhon, 
Salgado, and Marck \cite{BGSM93} (hereafter BGSM).  
In Appendix \ref{secApp:BGSM}, it is shown that 
the Bekenstein and Oron theory can also accommodate the 
BGSM formulation if a term involving $\Lie_k q^\alpha$ 
is kept in the MHD-Euler equation.

\subsection{Bekenstein-Oron 4-current}

From (\ref{eq:BOcur2}), the Bekenstein-Oron 4-current can be expressed as
\beqn
j^\alpha
&=& \frac1{\sqrt{-g}}\Lie_q(\rho u^\alpha \sqrt{-g})
\label{eq:cur2}
\\
&=& -\rho \Lie_u q^\alpha \,+\, u^\alpha  \na_\beta (\rho\, q^\beta).
\label{eq:cur3}
\eeqn
Let us recall that $j^\alpha$ is invariant under a change of $q^\alpha$
of the form (\ref{eq:change_q}).
Without loss of generality, a condition such as 
$q^\alpha u_\alpha = 0$, or $q^\alpha \na_\alpha t = 0$,
may be imposed, although these are not used below.

The 4-current must obey the helical symmetry, namely its Lie derivative along
$k^\alpha$ must vanish:
\beqn
\Lie_k j^\alpha 
\,=\, \na_\beta(\rho u^\alpha \Lie_k q^\beta\,-\,\rho u^\beta \Lie_k q^\alpha)
\,=\, 0 ,
\label{eq:LieBOcur}
\eeqn
where $\Lie_k q^\alpha \ne 0$.  
Using (\ref{eq:cur2}) and (\ref{eq:cur3}), we can write
\beqn
\Lie_k j^\alpha
&=& \frac1{\sqrt{-g}}\Lie_{[k,q]}(\rho u^\alpha \sqrt{-g})
\label{eq:Liecur2}
\\
&=& -\rho \Lie_u \Lie_k q^\alpha \,+\, u^\alpha  \na_\beta (\rho\, \Lie_k q^\beta)
\,=\, 0, 
\label{eq:Liecur3}
\eeqn
where the commutator notation $[k,q]^\alpha = \Lie_k q^\alpha$ is used.  Note 
the commutation relation $\Lie_k \Lie_u - \Lie_u \Lie_k = \Lie_{[k,u]}=0$, 
for $u^\alpha$ respects the symmetry.  In the above expressions for 
$\Lie_k j^\alpha$, it is noticed that we have again the freedom to add 
to $\Lie_k q^\alpha$ a vector proportional to $u^\alpha$, as 
$\Lie_k q^\alpha \mapsto
\Lie_k q^\alpha + \lambda u^\alpha$.

\subsection{Helically symmetric MHD-Euler equation}

We first rewrite the MHD-Euler equation 
by isolating the Lie derivative along the helical vector $k^\alpha$.
Using the decomposition (\ref{eq:4v_decomp}) of the 4-velocity, 
the MHD-Euler equation (\ref{eq:MHD-Euler_form1}) divided by $u^t$
is written
\beqn
(k^\beta+v^\beta) (dw)_\beal 
&=& 
- \na_\alpha(w_\beta k^\beta) 
\,+\, \Lie_k w_\alpha \,+\, v^\beta (dw)_\beal 
\nonumber\\
&=&0. 
\label{eq:MHD-Euler_form3}
\eeqn
Since $\eta_\alpha  u^\alpha = \Fabd u^\alpha q^\beta = 0$ for ideal MHD, 
we have 
\beq
  w_\alpha u^\alpha = (h u_\alpha + \eta_\alpha) u^\alpha = - h ,
\eeq
hence
\beq
w_\alpha k^\alpha 
\,=\, w_\alpha \left(\frac{u^\alpha}{u^t} - v^\alpha \right) 
\,=\, -\Big(\, \frac{h}{u^t} + w_\alpha v^\alpha \,\Big),
\eeq
Substituting this relation into (\ref{eq:MHD-Euler_form3}), we obtain 
\beq
\na_\alpha \Big(\,\frac{h}{u^t} + w_\beta v^\beta \,\Big) 
\,+\, \Lie_{k} w_\alpha \,+\, v^\beta (dw)_\beal \,=\, 0.  
\label{eq:MHD-Euler_form4}
\eeq

Since both $h u_\alpha$ and $F_{\alpha\beta}$ respect the helical symmetry, we
have, given the definition (\ref{eq:defw}) of $w_\alpha$, 
\beq
\Lie_k w_\alpha \,=\, \Lie_k (hu_\alpha \,+\, F_{\alpha\beta} q^\beta) 
\,=\, \Fabd \Lie_k q^\beta .  
\eeq
Hence Eq.~(\ref{eq:MHD-Euler_form4}) becomes
\beq
\na_\alpha \Big(\,\frac{h}{u^t} + w_\beta v^\beta \,\Big) 
\,+\, F_\albe \Lie_{k} q^\beta \,+\, v^\beta (dw)_\beal \,=\, 0.  
\label{eq:MHD-Euler_helical}
\eeq
Starting from this form of the MHD-Euler equation, let us discuss two 
cases: the corotational flow and the irrotational one. 
\paragraph{Corotational flow:}
The flow is \emph{corotational} if the fluid 4-velocity is parallel to the
Killing vector: $u^\alpha = u^t k^\alpha$.
This amounts to setting $v^\alpha =0$ in the decomposition (\ref{eq:4v_decomp})
of the 4-velocity. Accordingly, Eq.~(\ref{eq:MHD-Euler_helical}) reduces to 
\beq
\na_\alpha \Big(\,\frac{h}{u^t}\,\Big) 
\,+\, \Fabd \Lie_{k} q^\beta \,=\, 0.  
\label{eq:ME_corot}
\eeq
Note that, thanks to (\ref{eq:cur3}) 
and the ideal MHD condition (\ref{eq:idealMHD}), we have
\beq
\Fabd \Lie_k q^\beta \,=\, - \frac1{\rho u^t}\Fabd j^\beta
\label{eq:Liekq_j_comove}
\eeq
in the corotating case.

\paragraph{Irrotational flow:}

In the Bekenstein and Oron ideal MHD theory, the magnetized flow 
is called \emph{irrotational} when the vorticity $\omega_\albe=(d w)_\albe$ 
defined by (\ref{eq:def_vorticity}) vanishes identically.
The MHD-Euler equation (\ref{eq:MHD-Euler_form1})
is then always satisfied.
Via the Poincar\'e lemma, a flow is irrotational if, and only if, there exists (locally)
a potential $\Phi$ such that 
$w_\alpha = \na_\alpha\Phi$. 
Since $w_\alpha v^\alpha=v^\alpha \na_\alpha \Phi=\Lie_v \Phi$, 
and $v^\beta (dw)_\beal=0$, Eq.~(\ref{eq:MHD-Euler_helical}) reduces to 
\beq
\na_\alpha {\Big(} \frac{h}{u^t} + \Lie_v\Phi{\Big)}
\,+\,\Fabd \Lie_k q^\beta \,=\,0.
\label{eq:ME_irrot}
\eeq
Note that, contrary to the corotating case, the contribution of the Lorentz force
is divided into two terms: $\Fabd \Lie_k q^\beta$ and the term involving 
the potential $\Phi$.

\subsection{Integrability conditions}
\label{secApp:1stint}

Under the assumption of helical symmetry without any restriction 
on the fluid flow, the integrability condition for 
Eq.~(\ref{eq:MHD-Euler_helical}) is that the last two terms 
in the left hand side be the gradient of a function $f$, 
\beq
\Fabd \Lie_k q^\beta \,+\, v^\beta  (dw)_\beal 
\,=\, \na_\alpha f.
\label{eq:intcon_1}
\eeq
It may also be possible that each term is separately integrable, 
that is, with two functions $f$ and $g$, 
each term is a gradient, 
\beq
\Fabd \Lie_k q^\beta \,=\, -\Lie_k q^\beta (dA)_\beal \,=\, \na_\alpha f, 
\label{eq:intcon_Liekq}
\eeq
and 
\beq
v^\beta  (dw)_\beal \,=\, \na_\alpha g. 
\label{eq:intcon_vdw}
\eeq
Therefore, the problem of finding a current with which the helically 
reduced MHD-Euler equation has a first integral 
is replaced by the problem of finding the Lagrange multiplier 
$q^\alpha$ that satisfies the above integrability conditions.  
As mentioned in \cite{Bekenstein:2000sf}, however, 
the vector $q^\alpha$ is not a freely specifiable 
quantity, and hence 
it is not trivial to find such a $q^\alpha$, 
even for corotating or irrotational flow where 
the $v^\beta (dw)_\beal$ term vanishes and 
the integrability condition reduces to 
Eq.~(\ref{eq:intcon_Liekq}).

\section{Formulations for magnetized binary neutron stars 
in equilibrium}
\label{secApp:magfluid}

\subsection{Bekenstein and Oron's first integral for 
magnetized irrotational flow} 

As mentioned earlier, assuming the current is written as in 
Eq.~(\ref{eq:BOcur}), and the vector $q^\alpha$ respects the 
symmetry, the MHD-Euler equation is integrable 
for irrotational flow.  Since the canonical momentum $w_\alpha$ 
defined in Eq.~(\ref{eq:defw}) respects the symmetry, 
$\Lie_k w_\alpha =0$, 
and the velocity potential for the magnetized irrotational 
flow is defined by Eq.~(\ref{eq:irrot}), 
the first integral is written $\Lie_k \Phi = \rm constant$
(which is equivalent to $w_\alpha k^\alpha = \rm constant$), 
or more explicitly, from Eq.~(\ref{eq:ME_irrot}), 
\beq
\frac{h}{u^t} + \Lie_v\Phi \,=\, \cal E, 
\eeq
where $\cal E$ is a constant. 
Assuming a one-parameter EOS, 
we have three solvable equations, 
the normalization condition for the 4-velocity, 
the first integral, and the rest mass conservation equation, 
for the three variables $\{h, u^t, \Phi \}$.  The equation 
for $\Phi$ is derived in Sec.~\ref{sec:vpot_eq}.

\subsection{A first integral for 
 initial data of irrotational magnetized binaries} 

Since part of our motivation for calculating numerical 
solutions of compact binary systems is to prepare 
quasi-equilibrium solutions that can be used as 
initial data sets for binary inspiral simulations, 
we assume that the multiplier $q^\alpha$ can be specified freely 
on an initial spacelike hypersurface $\Sigma_t$.  
Then, when all fields and matter satisfy helical 
symmetry, and the vector $\Lie_k q^\alpha$ is, at least 
instantaneously, proportional to the helical killing vector, 
the term $\Fabd \Lie_k q^\beta$ becomes integrable 
\beq
\Lie_k q^\alpha = \Lie_k q^t\, k^\alpha,  
\label{eq:Liekq_propk}
\eeq
and the coefficient $\Lie_k q^t$ is a function of 
$A_\beta k^\beta$.  
Note that the assumption (\ref{eq:Liekq_propk}) is valid only 
for irrotational flow; for corotational flow 
$\Fabd u^\beta = 0$ implies 
$\Fabd \Lie_k q^\beta = \Lie_k q^t \Fabd k^\beta = 0$.  
From the Cartan identity (\ref{eq:Cartan}) and $\Lie_k A_\alpha = 0$, 
and the assumption (\ref{eq:Liekq_propk}), 
the term (\ref{eq:intcon_Liekq}) becomes
\beq
-\Lie_k q^\beta  (dA)_\beal 
\,=\,\Lie_k q^t\, \na_\alpha(A_\beta k^\beta).
\eeq
Hence, for irrotational flow, Eq.~(\ref{eq:ME_irrot}) is rewritten 
\beq
\na_\alpha{\Big(} \frac{h}{u^t} + \Lie_v\Phi{\Big)}
\,+\,\Lie_k q^t \, \na_\alpha(A_\beta k^\beta) \,=\,0, 
\label{eq:ME_irrot_dAk}
\eeq
and is integrable if there is a function $f$ such that
\beq
\Lie_k q^t \,=\, f(A_\beta k^\beta), 
\label{eq:ME_irrot_intcond}
\eeq
so that
\beq
\frac{h}{u^t} \,+\, \Lie_v\Phi
\,+\, \int \Lie_k q^t \, d(A_\beta k^\beta) \,=\,{\cal E},  
\label{eq:first_int}
\eeq
where ${\cal E}$ is a constant.  

If a data set on an initial hypersurface respects helical 
symmetry permanently, 
the current should necessarily be stationary, $\Lie_k j^\alpha = 0$. 
Substituting Eq.~(\ref{eq:Liekq_propk}) into Eq.~(\ref{eq:Liecur3}), 
we have 
\beq
\Lie_k j^\alpha
\,=\, - \rho k^\alpha \Lie_u \Lie_k q^t \,+\, \rho u^\alpha \Lie^2_k q^t\,=\,0, 
\label{eq:Liecurj_qpropk}
\eeq
where we have used the facts that $\rho$, or $u^\alpha$ respect 
the symmetry, and a relation $\na_\alpha k^\alpha = 0$.  
When the integrability condition (\ref{eq:ME_irrot_intcond}) 
is satisfied, a coefficient of $u^\alpha$ in Eq.~(\ref{eq:Liecurj_qpropk}) 
vanishes, $\Lie^2_k q^t=\Lie_k f(A_\alpha k^\alpha)=0$, and hence 
a sufficient condition for stationarity of the current 
$\Lie_k j^\alpha=0$ is that the coefficients of $k^\alpha$ in 
Eq.~(\ref{eq:Liecurj_qpropk}) vanish, 
\beq
\Lie_u \Lie_k q^t \,=\, \Lie_u f(A_\alpha k^\alpha) \,=\,0.  
\label{eq:Liekj_cond}
\eeq
This condition is equivalent to the component of 
the ideal MHD condition along $k^\alpha$, 
\beq
k^\alpha \Fabd u^\beta = -\Lie_u (A_\alpha k^\alpha) =0,
\eeq
and is rewritten, on the fluid support of $\Sigma_t$, as
\beq
\Lie_v (A_\alpha k^\alpha) =0, 
\label{eq:AKconstraint}
\eeq
that is, $A_\alpha k^\alpha$ is constant along the spatial 
velocity $v^\alpha$ defined by Eq.~(\ref{eq:4v_decomp}).  
However, as mentioned above, there is no guarantee that 
solutions calculated from the $q^\alpha$ of 
Eq.~(\ref{eq:ME_irrot_intcond}) satisfies 
Eq.~(\ref{eq:Liekj_cond}) or (\ref{eq:AKconstraint}).

As we choose $\Lie_k q^\alpha$ to be parallel to $k^\alpha$ 
in (\ref{eq:Liekq_propk}), we may further restrict $q^\alpha$ 
so that $q^\alpha$ itself is parallel to $k^\alpha$, 
\beq
q^\alpha = q^t\, k^\alpha.  
\label{eq:q_propk}
\eeq
We substitute (\ref{eq:q_propk}) to the current (\ref{eq:BOcur2}) 
to derive an explicit form for the current $j^\alpha$, 
\beqn
j^\alpha
&=& \Lie_{q^t k} (\rho u^\alpha) 
\,+\, \rho u^\alpha \na_\beta (q^t k^\beta)
\nonumber \\
&=& - \rho k^\alpha \Lie_u q^t \,+\, \rho u^\alpha \Lie_k q^t. 
\label{eq:curj_qpropk}
\eeqn
For example, 
\beq
\Lie_k q^t = {\rm constant} 
\label{eq:Liekqt}
\eeq
satisfies the stationarity of the current (\ref{eq:Liekj_cond}) 
and 
\beq
q^t \,=\, [\, at+b\phi+f_q(x^A) \,]k^\alpha 
\eeq
satisfies Eq.~(\ref{eq:Liekqt}), where $f_q$ is a 
function of coordinates $x^A$ $A=1,2$ orthogonal to 
$k^\alpha$, $k^\alpha \na_\alpha x^A=0$, 
and $a, b$ are parameters that satisfy 
\beq
a+b\Omega \,=\, 1.
\label{eq:Qpara}
\eeq
Remember that $t$ parametrizes the foliation and the symmetry 
vector is normalized as $k^\alpha \na_\alpha t=1$, and 
$\phi$ parametrizes circular orbits with parameter length $2\pi$ 
and $k^\alpha \na_\alpha \phi=\Omega$.

\subsection{A model with $q^\alpha = [\, at+b\phi+f_q(x^A) \,]\hq^\alpha$}

We next consider a more general form of $q^\alpha$ where neither 
$q^\alpha$ nor $\Lie_k q^\alpha$ is proportional to $k^\alpha$.  
Separating the dependence on the coordinate associated with the 
$k^\alpha$, we assume the form of the vector $q^\alpha$ to be 
\beq
q^\alpha \,=\, [\, at+b\phi+f_q(x^A) \,]\hq^\alpha ,
\label{eq:Qdef}
\eeq
where $\hq^\alpha$ respects the symmetry 
\beq
\Lie_k \hq^\alpha \,=\, 0, 
\label{eq:Qsym}
\eeq 
and hence the relation 
\beq
\Lie_k q^\alpha = \hq^\alpha
\label{eq:LieQdef}
\eeq
holds.  

For corotational or irrotational flows, 
the integrability condition (\ref{eq:intcon_Liekq}) 
is rewritten with the requirement that there exists a function $f$ such that
\beq 
\Fabd \hq^\beta \,=\, -\hq^\beta (dA)_\beal \,=\, \na_\alpha f, 
\label{eq:intcon_Q}
\eeq
or using the Cartan identity,  
\beq
\Lie_\hq A_\alpha = \na_\alpha (A_\beta \hq^\beta - f).
\eeq
When stationarity is imposed to the current, 
substituting Eq.~(\ref{eq:LieQdef}) to Eq.~(\ref{eq:LieBOcur}), 
we have 
\beq
\Lie_k j^\alpha 
\,=\, \na_\beta(\rho u^\alpha \hq^\beta\,-\,\rho u^\beta \hq^\alpha)
\,=\, 0. 
\label{eq:LieBOcurQ}
\eeq
Then, from Eq.~(\ref{eq:Qdef}) and Eq.~(\ref{eq:BOcur}), 
the current becomes 
\beq
j^\alpha
\,=\,
(\rho u^\alpha \hq^\beta\,-\,\rho u^\beta \hq^\alpha)\na_\beta [\, at+b\phi+f_q(x^A) \,].  
\label{eq:BOcurQ}
\eeq

\paragraph{Corotating flow: }
This model can be applied to corotating flow, as long as 
one can find a particular form of $\hq^\alpha$ that satisfies 
Eq.~(\ref{eq:intcon_Q}) as well as the stationarity and ideal MHD 
conditions consistently.  
For corotating flow, $u^\alpha = u^t k^\alpha$, 
Eq.~(\ref{eq:BOcurQ}) becomes
\beq
j^\alpha
\,=\,
\rho u^t k^\alpha \hq^\beta \na_\beta [\, at+b\phi+f_q(x^A) \,]
\,-\,\rho u^t \hq^\alpha .
\label{eq:BOcurQ_corot}
\eeq
Assuming $f_q(x^A)=0$ and 
using $a+\Omega b=1$, the combination of $t$ and $\phi$ components 
$j^\phi - \Omega j^t$ becomes 
\beqn
j^\phi - \Omega j^t 
&=& \rho u^t(k^\alpha \hq^\beta\,-\, k^\beta \hq^\alpha) \na_\alpha \phi \na_\beta t
\nonumber\\
&=& -\rho u^t(\hq^\phi\,-\, \Omega \hq^t).  
\eeqn
As discussed in Appendix \ref{secApp:BGSM}, 
when the system is stationary and axisymmetric, 
and if $\hq^\alpha$ satisfies 
\beq
\hq^\alpha \,=\, f(A_\phi) \phi^\alpha, 
\eeq
the formulation becomes the same as that of \cite{BGSM93} 
for a magnetized rotating neutron star.

\paragraph{A trivial model for the irrotational flow: }
When $\hq^\alpha$ is taken to be parallel to $k^\alpha$,  
\beq
\hq^\alpha = \hq^t k^\alpha, 
\eeq
with $\Lie_k \hq^t = 0$, 
the first integral is derived as in the previous section, 
if $\hq^t$ is a function of $A_\alpha k^\alpha$; 
Eq.~(\ref{eq:intcon_Q}) becomes 
\beq
-\hq^\beta (dA)_\beal = \hq^t \na_\alpha (A_\beta k^\beta) = \na_\alpha f.  
\label{eq:intcon_Q2}
\eeq
The current (\ref{eq:BOcurQ}) in this case is written 
\beq
j^\alpha
\,=\,
\rho u^t \{v^\alpha 
- k^\alpha v^\beta\na_\beta [\, b\phi+f_q(x^A) \,]\}\,\hq^t.  
\label{eq:BOcurhq}
\eeq
A trivial solution to the condition (\ref{eq:intcon_Q2}) is 
\beq
\hq^t = \mbox{constant}.  
\eeq

\subsection{Equation for the velocity potential $\Phi$}
\label{sec:vpot_eq}

To write down an equation for the velocity potential $\Phi$ 
for magnetized irrotational flow used in an actual numerical code, 
we introduce a 3+1 decomposition of the spacetime.  
In this section, spatial indices are Latin.  
The spacetime ${\cal M}=\mathbb{R}\times \Sigma$ is foliated 
by a family of spacelike 
hypersurfaces $(\Sigma_t)_{t\in\mathbb{R}}$
parametrized by $t$.
The future-pointing unit normal to the hypersurface $\Sigma_t$ 
is defined by $n_\alpha = -\alpha \na_\alpha t$, where $\alpha$
is the lapse function.  Then the generator of time translations 
in an inertial frame $t^\alpha$, and rotating frame (helical vector) 
$k^\alpha$ are related to $n^\alpha$ by 
$t^\alpha = \alpha n^\alpha+\beta^\alpha$ and 
$k^\alpha = \alpha n^\alpha+ \omega^\alpha$ respectively, where 
$\beta^\alpha$ and $\omega^\alpha$ denote a spatial shift vector 
in each frame, and are related by 
$\omega^\alpha = \beta^\alpha + \Omega \phi^\alpha$.  
The spatial metric $\gmabd(t)$ induced on $\Sigma_t$ 
by the spacetime metric $g_{\alpha\beta}$
is equal to the projection tensor orthogonal to $n^\alpha$, 
$\gamma_{\albe} = \gabd+n_\alpha n_\beta$, restricted to $\Sigma_t$.  
In a chart $(t,x^a)$, the metric $\gabd$ has the form
\beq 
ds^2 
= 
-\alpha^2dt^2+ \gamma_{ab}(dx^a+\beta^a dt)(dx^b+\beta^b dt). 
\eeq
The covariant derivative associated with the spatial metric 
$\gmabd$ is denoted by $D_a$.  

In the formulation for irrotational flow, 
the number of independent variables becomes 
three \cite{irbns_formulation,UryuCFWL}.  
As independent variables, we choose 
the relativistic enthalpy per baryon mass, 
the time component of the 4-velocity, and the 
velocity potential, $\{h,u^t,\Phi\}$.  
For the first two variables, 
the first integral Eq.~(\ref{eq:ME_irrot_dAk}) and 
the normalization of the 
4-velocity $u_\alpha u^\alpha=-1$ are solved.  
Using a relation derived from Eqs.~(\ref{eq:4v_decomp}) 
and (\ref{eq:irrot}), 
\beq
v_a + \omega_a = \frac1{hu^t}(D_a\Phi - \eta_a), 
\label{eq:spatialvpot}
\eeq
these equations are rewritten, 
\beqn
&& 
\frac{h}{u^t}
\,+\, v^a D_a \Phi
\,+\, \int \Lie_k q^t\,d(A_\alpha k^\alpha)
\,=\,{\cal E}, 
\label{eq:hoverut}
\\
&&
h^2\left[(\alpha u^t)^2 - 1 \right] 
\,=\, (D^a\Phi - \eta^a)(D_a\Phi - \eta_a), 
\label{eq:ut}
\eeqn
where $\eta_a$ is a spatial projection of 
$\eta_\alpha$, $\eta_a = \gamma_a{}^\alpha \eta_\alpha$.  

An equation to calculate the velocity potential $\Phi$ is derived 
from the rest mass conservation law, Eq.~(\ref{eq:consrhos}), 
\beqn
\frac{1}{\sqrt{-g}}\Lie_u (\rho\sqrt{-g})
&=& \frac{1}{\alpha\sqrt{\gamma}}\Lie_v (\rho u^t \alpha\sqrt{\gamma}) 
\nonumber\\
&=& \frac{1}{\alpha}D_a(\alpha \rho  u^t v^a) =0.  
\eeqn
Substituting Eq.~(\ref{eq:spatialvpot}) in the above relation, 
we have an elliptic equation for $\Phi$, 
\beq
D^a D_a\Phi = D_a(\eta^a + hu^t\omega^a)
- (D_a\Phi - \eta_a - hu^t\omega_a)\frac{h}{\alpha\rho}D^a\frac{\alpha\rho}{h}.
\label{eq:LapPhi}
\eeq
This equation is solved with a Neumann boundary condition 
to impose the fluid 4-velocity $u^\alpha$ to follow 
the surface of the star.
The surface is defined by the vanishing pressure $p=0$, where 
the relativistic enthalpy is chosen to be $h=1$ which is always 
possible when a one-parameter equation of state is assumed.  
Hence, the boundary condition is written 
\beq
u^\alpha \na_\alpha h =0  \ \ \mbox{ at }\ \  h=1.  
\eeq
and, using $\Lie_k h = 0$ and Eq.~(\ref{eq:spatialvpot}), 
it is rewritten, 
\beq
(D^a\Phi - \eta^a - hu^t\omega^a)D_a h =0.
\eeq
where $\na_\alpha h$ and $D_a h$ are normal to the stellar surface.

\section{Discussion}

\subsection{First law associated with the Bekenstein and Oron 
Lagrangian}

The Lagrangian density of the Bekenstein and Oron 
ideal MHD theory \cite{Bekenstein:2000sf} 
is based on Schutz's Lagrangian density 
for relativistic fluids \cite{Schutz:1970my}.  Our 
Lagrangian density for a relativistic fluid 
$\Lagm = -\epsilon\sqrt{-g}$ (\ref{eq:Lagm}), and 
the Lagrangian variation applied to it, is equivalent for the purpose of deriving the first law.  
Then, we rewrite the Lagrangian corresponding to 
that of Bekenstein and Oron as 
\beq
\Lag \,=\, 
\left(\frac1{16\pi} R \,-\, \epsilon 
\,-\,\frac1{16\pi} \Fabd \Fabu \,+\, \Fabd \rho u^\alpha q^\beta 
\right)\sqrt{-g}, 
\label{eq:LagBO}
\eeq
in which the interaction term is replaced by a term $\Fabd  u^\alpha$ times the Lagrange multiplier $\rho q^\alpha$ which enforces
the ideal MHD condition $\Fabd u^\alpha=0$. 

Associating this Lagrangian with the charge $Q$ (\ref{eq:Q}), 
we can derive the first law; a calculation of the variation 
$\dl Q$ is shown in Appendix \ref{secApp:dlQBO}.  
Now, the derived first law is for the ideal MHD flow, while our 
first law (\ref{eq:1stlaw_org}) is valid for more general 
MHD flows.  Obviously, the argument in Sec.~\ref{sec:conserv} 
applies to the case with the Lagrangian (\ref{eq:LagBO}).  
Hence, 
if a sequence of magnetized binary solutions in equilibrium 
is constructed assuming 
conservation of rest mass, entropy, magnetized 
circulation, magnetic flux, black hole 
surface area and charge for a black hole - neutron star 
binary, the first law in the form 
$\dl Q=0$, or $\dl \Madm = \Omega \dl J$
for asymptotically flat systems, is satisfied 
as for non-magnetized ones, and for the latter 
case, one can apply a turning point theorem to 
locate a point where the stability of solution 
changes \cite{turning1}.

\subsection{First integral of MHD-Euler equation }

As mentioned in Sec.~\ref{sec:integrability}, 
a first integral of the relativistic MHD-Euler equation 
is almost crucial for developing a successful method 
to compute equilibrium binary solutions numerically.  
When we derive a first integral, we need to specify 
a form of the vector $q^\alpha$, which should be consistent 
with the stationarity as well as the ideal MHD condition.  
However, since $q^\alpha$ is not a freely specifiable vector, 
it is not guaranteed that a set of equations admit such a 
$q^\alpha$ as solution in general.  
Also a difficulty to have a helically symmetric irrotational 
binary solution in ideal MHD may be explained physically 
as follows.  Because the magnetic flux is frozen into the fluid for ideal MHD, 
when the binary system is seen 
in the rotating frame, a poloidal component of the magnetic field 
may be winded up, since the neutron star is spinning 
in this frame.  This argument does not rule out the 
possibility to have a helically symmetric magnetized 
binary neutron stars, although it is not trivial at all 
to find a $q^\alpha$ that gives such solutions.  

In Sec.~\ref{secApp:magfluid}, we discuss a formulation 
for computing equilibrium solutions of magnetized binary neutron stars
and a possible candidate for a first integral of the relativistic 
MHD-Euler equation in ideal MHD flows.  
Our proposal is to assume $\Lie_k q^\alpha$ be proportional 
to the helical vector $k^\alpha$.  
It could be possible that this condition is violated as 
the solution is evolved in time, that is, a solution 
calculated from the first integral in the Appendix might not 
respect the helical symmetry or the ideal MHD condition.  
It would be applicable, however, 
for computing initial data for merger simulations 
of magnetized compact objects, because it may be allowed 
to freely specify $\Lie_k q^\alpha$, at least instantaneously 
on an initial hypersurface.

In Sec.~\ref{secApp:magfluid} we also write down a set of equations 
to be solved for an equilibrium of irrotational neutron star 
in a binary system.  
The formulation for solving
the Einstein and Maxwell equations are not presented in this paper.  
In usual ideal MHD simulations, 
the electric current $j^\alpha$ does not contain dynamical 
degrees of freedom and, accordingly, the Maxwell equation 
becomes an evolution equation for the magnetic flux density.  
This equation is again hard to integrate when the stationarity 
condition is imposed.  Therefore our plan is to choose 
the electromagnetic potential one form $A_\alpha$ as 
a variable and to write the Maxwell equation as a set of 
elliptic equations.  
These elliptic equations can be solved 
with the same numerical method we have developed to 
solve for the metric potentials of gravitational fields 
\cite{YBRUF06,UryuCFWL}.  
Our next project is to develop such a numerical code.

\acknowledgments
We thank Brandon Carter, John L.\ Friedman and Ichiro Oda 
for enlightening discussions and suggestions.  
This work was supported by 
JSPS Grant-in-Aid for Scientific Research(C) 20540275, 
MEXT Grant-in-Aid for Scientific Research
on Innovative Area 20105004, 
NSF grants No.~PHY0071044 and PHY0503366, 
NASA grant No.~NNG05GB99G, 
and ANR grant 06-2-134423 \emph{M\'ethodes 
math\'ematiques pour la relativit\'e g\'en\'erale}.  
CM thanks the Greek State Scholarships Foundation for support during 
the early stages of this work and the Paris Observatory and 
the University of Wisconsin-Milwaukee for travel support.
KU and EG acknowledge support from the JSPS Invitation Fellowship 
for Research in Japan (Short-term) and the invitation program 
of foreign researchers at the Paris Observatory.

\appendix

\section{Variation of the Lagrangian} 
\label{s:Variation_Lagrangian}

We begin with a classical action for an Einstein-Maxwell theory coupled 
with a perfect fluid carrying electric current, 
\beq
S = \int \Lag \,d^4x, 
\eeq
\beqn
\Lag &=& \LagG \,+\,\Lagm \,+\,\LagF \,+\,\LagI
\nonumber \\
 &=& 
\left(\frac1{16\pi} R \,-\, \epsilon 
\,-\,\frac1{16\pi} \Fabd \Fabu \,+\, A_\alpha j^\alpha
\right)\sqrt{-g}.
\nonumber\\
\label{eq:Lag}
\eeqn
We first define the Lagrange perturbation for the fluid.  

\subsection{Lagrange displacement}
\label{secApp:Lag}

We describe a perfect fluid by its four-velocity $u^\alpha$ and 
stress-energy tensor 
\beq 
        T^{\alpha\beta} = \epsilon u^\alpha u^\beta + p q^{\alpha\beta}, 
\eeq     
where $p$ is the fluid's pressure, $\epsilon$ its energy density, and  
\beq  
        q^{\alpha\beta} = g^{\alpha\beta} + u^\alpha u^\beta 
\eeq 
is the projection tensor orthogonal to $u^\alpha$. We assume that the fluid 
satisfies an equation of state of the form 
\beq  
         p = p(\rho,s),\    \epsilon = \epsilon(\rho,s),   
\eeq   
with $\rho$ the baryon-mass density and $s$ the entropy per unit baryon
mass.  
(That is, $\rho := m_B n$, with $n$ the number density of baryons 
and $m_B$ the mean baryon mass.)  

The electromagnetic stress-energy tensor is given by 
\beq
\TabFu\,=\,
\frac1{4\pi}\left(F^{\alpha\gamma}F^\beta{}_\gamma
-\frac14\gabu\Fcdd\Fcdu\right), 
\label{eqApp:TabFu}
\eeq
where electromagnetic field 2-form $\Fabd$ relates to 
the potential 1-form by 
\beq
\Fabd = (dA)_\albe = 
\na_\alpha A_\beta-\na_\beta A_\alpha.  
\eeq

Given a family of magnetized perfect-fluid Einstein-Maxwell 
spacetimes specified by 
\beq 
{\cal Q}(\lambda) := [g_{\alpha\beta}(\lambda), u^\alpha(\lambda),  
\rho(\lambda), s(\lambda), A_\alpha(\lambda), j^\alpha(\lambda)],   
\eeq 
one defines the Eulerian change in each quantity by 
$\delta {\cal Q} := \frac d{d\lambda} {\cal Q}(\lambda).$ 

We introduce a Lagrangian displacement 
$\xi^\alpha$ in the following way: Let ${\cal Q} := {\cal Q}(\lambda)$, and let 
$\Psi_\lambda$ be a diffeomorphism mapping each trajectory (worldline) of the
initial fluid to a corresponding trajectory of the configuration 
${\cal Q}(\lambda)$.  Then the tangent $\xi^\alpha(P)$ to the path 
$\lambda \rightarrow \Psi_\lambda(P)$ can be regarded as a vector 
joining the fluid element at $P$ in the configuration ${\cal Q}(\lambda)$ to a 
fluid element in a nearby configuration. The Lagrangian change 
in a quantity at $\lambda=0$, is then given by 
\beq 
        \Delta  {\cal Q} := \frac d{d\lambda} \Psi_{-\lambda}{\cal Q} 
        (\lambda)|_{\lambda = 0}   
          = (\delta + \Lie_\xi ){\cal Q}. 
\eeq

The fact that $\Psi_\lambda$ maps fluid trajectories to fluid trajectories 
and the normalization $u^\alpha u_\alpha=-1$ imply  
\beq 
        \Dl u^\alpha = \frac{1}{2}u^\alpha u^\beta u^\gamma 
        \Delta g_{\beta\gamma}.
\label{eq:Dlu}   
\eeq

\subsection{Variation of Lagrangian}

Although the variation of the Lagrangian density (\ref{eq:Lag}) 
is well known, those calculations are summarized 
below to clarify notation and conventions. 
A surface term $\Theta^\alpha$ is kept for 
the calculation of the first law in Sec.~\ref{sec:1stlaw}.

\paragraph*{The variation of the Einstein-Hilbert Lagrangian}
is written as 
\beq
\frac1{\sqrt{-g}} \dl \LagG 
\,=\, -\frac1{16\pi}\Gabu \dl \gabd \,+\, \na_\alpha \ThGau
\eeq
\beq 
\ThGau
\,=\, \frac{1}{16\pi}(g^{\alpha\gamma}g^{\beta\delta} 
                     -g^{\alpha\beta}g^{\gamma\delta}) 
                     \na_\beta\dl g_{\gamma\delta}.   
\eeq 

\paragraph*{The variation of the perfect fluid Lagrangian} 
is described by the Lagrange perturbations.  Considering
general perturbations in which the entropy and baryon mass of each fluid 
element are not conserved along the family ${\cal Q}(\lambda)$, one obtains 
\beq 
\frac{\Dl\rho}{\rho} 
\,=\, -\frac1{\rho\sqrt{-g}}u_\alpha \Dl(\rho u^\alpha \sqrt{-g})
\,-\, \frac{1}{2}\qabu\Dlgab;  
\label{drho}
\eeq 
and the local first law of thermodynamics for the fluid, 
\beq 
        \Dl\epsilon = \rho T \Dl s + h \Dl\rho,  
\eeq 
with the relativistic enthalpy $h$ defined by 
\beq 
        h = \frac{\epsilon+p}{\rho},  
\eeq 
yields 
\beq 
\frac{\Dl\epsilon}{\epsilon+p}
\,=\, \frac{\rho T}{\epsilon+p}\Dl s \,+\, \frac{\Dl\rho}{\rho}.  
\eeq
Hence, we have 
\beq 
\Dl\epsilon \,=\, \rho T \Dl s 
\,-\, \frac1{\sqrt{-g}}h u_\alpha\Dl(\rho u^\alpha\sqrt{-g})
\,-\, \frac12(\epsilon+p)\qabu\Dl\gabd.  
\eeq

From these relations, the variation of the Lagrangian density 
for a perfect fluid 
\beq 
        \Lagm =  - \epsilon \sqrt{-g} 
\label{eq:Lagm}
\eeq 
becomes
\beqn 
\frac1{\sqrt{-g}}\dl\Lagm 
&=& -\frac1{\sqrt{-g}}\dl(\epsilon \sqrt{-g}) 
\nonumber \\  
&=& -\frac1{\sqrt{-g}}\Dl(\epsilon \sqrt{-g}) 
\,+\,\frac1{\sqrt{-g}}\Lie_\xi(\epsilon \sqrt{-g})
\nonumber \\  
&=& - \Dl\epsilon
\,-\, \epsilon\, \frac12\,\gabu\Dl\gabd
\,+\, \na_\alpha(\epsilon \xi^\alpha)
\nonumber \\  
&=& -\rho T \Dl s 
\,+\, \frac{1}{\sqrt{-g}} hu_\alpha\Dl(\rho u^\alpha \sqrt{-g})
\nonumber \\
&&+\  \frac12\,\Tabu\dl\gabd
\,-\, \xi_\alpha\na_\beta\Tabu
\,+\, \na_\alpha \Thmau, 
\nonumber \\
\label{eq:dlLagM}
\eeqn 
with the surface term 
\beq 
\Thmau\,=\, (\epsilon+p)\qabu\xi_\beta .
\eeq

\paragraph*{The variation of the Lagrangian for the electromagnetic field} 
\beq 
        \LagM = \,-\,\frac1{16\pi} \Fabd \Fabu \sqrt{-g},  
\label{eq:LagM}
\eeq 
is calculated as 
\beqn
\frac1{\sqrt{-g}}\dl \LagM
&=&
-\frac1{16\pi\sqrt{-g}}\dl(\Fabd\Fabu\sqrt{-g})
\nonumber\\
&=&
-\frac1{16\pi}\left[\,
2(d\,\dl A)_\albe \Fabu
\,+\,2F_{\alpha\gamma}F_\beta{}^\gamma\,\dl\gabu
\phantom{\frac12}
\right.
\nonumber\\
&&
\left.\qquad
\,+\,2\Fcdd\Fcdu\,\frac12\,\gabu\dl\gabd
\,\right]
\nonumber\\
&=&
\frac12\,\TabMu \dl\gabd
\,-\,\frac1{4\pi}\na_\beta\Fabu\dl A_\alpha
\,+\,\na_\alpha \ThMau, 
\nonumber\\
\eeqn
where $\ThMau$ is defined by
\beq
\ThMau\,=\,\frac1{4\pi}F^{\beta\alpha}\dl A_\beta.
\eeq

\paragraph*{The variation of the interaction term}
between matter and the electromagnetic field, 
\beq
\LagI\,=\,A_\alpha j^\alpha \sqrt{-g},
\eeq
becomes
\beqn
\frac1{\sqrt{-g}}\dl\LagI
&=&
\dl A_\alpha j^\alpha 
\,+\, A_\alpha \frac1{\sqrt{-g}}\Dl (j^\alpha \sqrt{-g}) 
\nonumber\\
&&
\,-\, A_\alpha \frac1{\sqrt{-g}}\Lie_\xi (j^\alpha \sqrt{-g}). 
\eeqn
Using the relation 
\beq
\frac1{\sqrt{-g}}\Lie_\xi (j^\alpha \sqrt{-g})
\,=\,
\na_\beta(j^\alpha \xi^\beta  - j^\beta \xi^\alpha)
\,+\, \xi^\alpha \na_\beta j^\beta, 
\eeq
we have 
\beqn
\frac1{\sqrt{-g}}\dl\LagI
&=&
j^\alpha \dl A_\alpha 
\,+\, A_\alpha \frac1{\sqrt{-g}}\Dl (j^\alpha \sqrt{-g}) 
\nonumber\\
&+& \xi^\alpha \left[\,\Fabd\, j^\beta 
\,-\, A_\alpha \na_\beta j^\beta\,\right]
\,+\, \na_\alpha \ThIau, 
\qquad
\label{eq:dlLagI}
\eeqn
where the surface term is defined by 
\beq
\ThIau \,=\,
A_\beta (j^\alpha \xi^\beta  - j^\beta \xi^\alpha).
\eeq

\paragraph*{Variation of the Lagrangian density:} 
Finally, the above terms are collected and 
the variation of the Lagrangian density (\ref{eq:Lag})
is derived, 
\beqn
\frac1{\sqrt{-g}}\dl\Lag 
&=& \frac1{\sqrt{-g}}(\dl\LagG \,+\,\dl\Lagm \,+\,\dl\LagM \,+\,\dl\LagI)
\nonumber \\
&=& -\rho T \Dl s 
\,+\, \frac{1}{\sqrt{-g}} hu_\alpha\Dl(\rho u^\alpha \sqrt{-g})
\nonumber \\
&&+\ A_\alpha \frac1{\sqrt{-g}}\Dl (j^\alpha \sqrt{-g}) 
\nonumber \\
&&-\ \frac1{16\pi}\left[\,\Gabu -8\pi(\Tabu+\TabMu)\,\right]\dl \gabd
\nonumber \\
&&-\ \frac1{4\pi}(\na_\beta\Fabu - 4\pi j^\alpha)\dl A_\alpha
\nonumber \\
&&-\ \xi^\alpha\left[\,\na_\beta\Tba - \Fabd j^\beta
\,+\, A_\alpha \na_\beta j^\beta \,\right]
\nonumber \\
&&
+\ \na_\alpha \Thau, 
\label{eq:dlLag}
\eeqn
where the surface term $\Thau$ is defined by
\beqn
\Thau 
&=& \ThGau \,+\,\Thmau\,+\,\ThMau\,+\,\ThIau
\nonumber \\
&=& \frac{1}{16\pi}(g^{\alpha\gamma}g^{\beta\delta} 
                     -g^{\alpha\beta}g^{\gamma\delta}) 
                     \na_\beta\dl g_{\gamma\delta}
\,+\, \frac1{4\pi}F^{\beta\alpha}\dl A_\beta
\nonumber\\
&&
+\ (\epsilon+p)\,\qabu\xi_\beta 
\,+\, A_\beta (j^\alpha \xi^\beta  - j^\beta \xi^\alpha).
        \label{eq:def_Theta}
\eeqn

\section{Calculation of $\dl (Q - \sum\limits_i Q_i)$}
\label{secApp:dlQ}

In calculating a contribution from the volume integral 
to the charge (\ref{eq:dlQ_vol}), we restrict 
the gauge in two ways:  We use the 
diffeomorphism gauge freedom to set $\delta k^\alpha =0$.  
The description of fluid perturbations in terms of a Lagrangian 
displacement $\xi^\alpha$ has a second 
kind of gauge freedom: a class of trivial displacements, including 
all displacements of the form $f u^\alpha$, yield no Eulerian 
change in the fluid variables.  We use this freedom to 
set $\Delta t =0$.  Because $\delta t =0$ ($t$ is not dynamical), 
this is equivalent to the condition $\xi^t = 0$.  
The relation (\ref{eq:Dlu}) now implies    
\beq
\frac{\Delta u^t}{u^t} = \frac12 u^\alpha u^\beta 
        \Delta g_{\alpha\beta}. 
\label{eq:dgamma}
\eeq
Then, from Eqs. (\ref{eq:Dlu}) and (\ref{eq:dgamma}), we have 
$\Delta u^\alpha = \Delta u^t (k^\alpha + v^\alpha)$, while, by 
$u^\alpha = u^t(k^\alpha + v^\alpha)$,  
$\Delta u^\alpha = \Delta [ u^t (k^\alpha + v^\alpha)];$
thus
\beq 
\Delta (k^\alpha + v^\alpha) = 0.
\label{eq:dv}  
\eeq
Then, in the variation of the Lagrangian density 
(\ref{eq:dlLag}), a term involving a perturbation 
of the rest mass density is rewritten 
\beq
hu_\alpha \Dl(\rho u^\alpha \sqrt{-g})
\,=\, -\,\frac{h}{u^t} \Dl(\rho u^t \sqrt{-g}) .
\eeq

        To find the change $\delta Q$ in the Noether charge, 
we first compute the difference, 
\beq
\delta \Big(Q - \sum_i Q_i\Big), 
\eeq 
between the charge on the sphere $S$ and the sum of the charges 
on the black holes ${\cal B}_i$.  

The difference in the Komar charge
Eq.~(\ref{eq:komar})
is associated with the Lagrangian density as 
\beqn
&& 
Q_K -\sum_i Q_{Ki} 
\nonumber\\
&=&
-\, \int_\Sigma \left(\frac{1}{16\pi} R 
\,-\,\epsilon \,-\,\frac1{16\pi}\Fabd\Fabu
\,+\,A_\alpha j^\alpha \right)k^\gamma dS_\gamma
\nonumber\\
&&
-\, \int_\Sigma \,
\left(\,\Tab + \TabF \,\right) k^\beta dS_\alpha 
\nonumber\\
&&
-\, \int_\Sigma 
\left(\epsilon \,+\,\frac1{16\pi}\Fabd\Fabu
\,-\,A_\alpha j^\alpha \right) k^\gamma dS_\gamma 
\nonumber\\
&&
-\, \frac{1}{8\pi} \int_\Sigma 
\left[\, \Gab \,-\, 8\pi (\Tab + \TabF)\,\right] k^\beta dS_\alpha .
\label{eq:qk_Aj}
\eeqn
Using the relations
\beqn 
        -\Tab k^\beta\dSa 
        &=& - \Tab(k^\beta+v^\beta)\dSa + \Tab v^\beta\dSa 
\nonumber \\ 
        &=& \epsilon\,k^\alpha \dSa 
        + (\epsilon + p) u^\alpha u_\beta v^\beta \dSa, 
\eeqn 
and
\beqn
&&\!\!\!\!\!\!\!\!\!\!\!\!\!\!
-\, \TabF \,k^\beta dS_\alpha 
-\, \left(\frac1{16\pi}\Fabd\Fabu
\,+\,A_\alpha j^\alpha \right)k^\gamma dS_\gamma
\nonumber\\
&=&
-\,\frac1{4\pi} F^{\alpha\gamma}
\left[\,\Lie_k A_\gamma - \na_\gamma(k^\beta A_\beta)\,\right]dS_\alpha
\,+\, A_\alpha j^\alpha k^\gamma dS_\gamma 
\nonumber\\
&=&
-\,\frac1{4\pi} F^{\alpha\gamma} \Lie_k A_\gamma dS_\alpha
\,+\,\frac1{4\pi} \na_\gamma (F^{\alpha\gamma}k^\beta A_\beta) dS_\alpha
\nonumber\\
&&
-\,\frac1{4\pi} k^\gamma A_\gamma 
(\na_\beta \Fabu - 4\pi j^\alpha) dS_\alpha
\nonumber\\
&&
\,+\,
A_\alpha(j^\alpha k^\gamma - j^\gamma k^\alpha) dS_\gamma ,
\label{eq:FFFq_Aj}
\eeqn
Eq.~(\ref{eq:qk_Aj}) is rewritten 
\beqn
&&
 Q_K -\sum_i Q_{Ki} 
\nonumber\\
&=&
-\, \int_\Sigma \Lag \,d^3x
\,+\,\int_\Sigma (\epsilon + p)u^\alpha u_\beta v^\beta dS_\alpha 
\nonumber\\
&&
\,-\,\frac1{4\pi}\int_\Sigma  F^{\alpha\gamma} \Lie_k A_\gamma dS_\alpha 
\,+\,\frac1{4\pi} \int_{\pa\Sigma}
k^\gamma A_\gamma F^{\alpha\beta} dS_{\alpha\beta}
\nonumber\\
&&
\,+\, \int_\Sigma 
A_\beta(j^\beta k^\alpha - j^\alpha k^\beta) dS_\alpha 
\nonumber\\
&&
\,-\, \frac{1}{8\pi} \int_\Sigma 
\left[\, \Gab \,-\, 8\pi (\Tab + \TabF)\,\right] k^\beta dS_\alpha 
\nonumber\\
&&
-\,\frac1{4\pi} \int_\Sigma k^\gamma A_\gamma 
(\na_\beta \Fabu - 4\pi j^\alpha) dS_\alpha .
\label{eq:QK_Aj}
\eeqn

The variation of Eq.~(\ref{eq:QK_Aj}) is then 
\beqn
&&
\delta \Big(Q_K -\sum_i Q_{Ki}\Big)
\nonumber\\
&=& \,-\, \int_\Sigma \dl {\cal L}\, d^3x 
\,+\, \int_\Sigma \Dl\left[(\epsilon + p) u^\alpha u_\beta v^\beta \dSa\right] 
\nonumber\\
&&
\,-\,\frac1{4\pi}\, \dl\int_\Sigma  F^{\alpha\gamma} \Lie_k A_\gamma dS_\alpha 
\,+\,\frac1{4\pi}\, \dl\int_{\pa\Sigma}
k^\gamma A_\gamma F^{\alpha\beta} dS_{\alpha\beta}
\nonumber\\
&&
\,+\, \int_\Sigma 
\Dl\left[A_\beta(j^\beta k^\alpha - j^\alpha k^\beta) \dSa \right]
\nonumber\\
&&
\,-\, \frac{1}{8\pi}\,\dl \int_\Sigma 
\left[\, \Gab \,-\, 8\pi (\Tab + \TabF)\,\right] k^\beta dS_\alpha, 
\nonumber\\
&&
-\,\frac1{4\pi}\,\dl \int_\Sigma k^\gamma A_\gamma 
(\na_\beta \Fabu - 4\pi j^\alpha) dS_\alpha .  
\label{eq:dlQK_Aj}
\eeqn 
The integrand of the second term becomes
\beqn
&&
\Dl\left[(\epsilon + p) u^\alpha u_\beta v^\beta \dSa\right] 
\nonumber\\
&=&
h u_\beta v^\beta \,\Dl(\rho u^\alpha   \dSa) 
\,+\,
v^\beta \,\Dl(h u_\beta )\rho u^\alpha \, \dSa 
\nonumber\\
&&
\,+\,
(\epsilon + p) u^\alpha u^\beta \,\Lie_k \xi^\beta \, \dSa , 
\label{eq:Dlhuv}
\eeqn
where $\Dl v^\beta  = -\Dl k^\beta = \Lie_k \xi^\beta$ was used, 
and the integrand of the fifth term is 
\beqn
&&\!\!\!\!\!\!\!\!\!\!\!\!\!\!\!
\Dl\left[A_\beta(j^\beta k^\alpha - j^\alpha k^\beta) \dSa \right]
\nonumber\\
&=&
\Dl A_\beta\,(j^\beta k^\alpha - j^\alpha k^\beta)\, \dSa
\nonumber\\
&& 
\,+\,A_\beta \frac1{\sqrt{-g}}\Dl (j^\beta \sqrt{-g})k^\alpha \dSa
\nonumber\\
&& 
\,-\, A_\beta k^\beta\, \Dl (j^\alpha \dSa)
\nonumber\\
&&
\,+\, A_\beta\,(j^\alpha \Lie_k \xi^\beta - j^\beta \Lie_k \xi^\alpha)\, \dSa
\label{eq:DlAj}
\eeqn
where $\Dl k^\alpha = \Lie_\xi k^\alpha = - \Lie_k \xi^\alpha$, 
because of our gauge choice $\dl k^\alpha = 0$.

The variation of $Q_L -\sum\limits_i Q_{Li}$ is given by
\beqn 
&&\!\!\!\!\!\!\!\!\!\!\!\!\!\!\!
\delta \Big( Q_L -\sum_i Q_{Li}\Big)
\nonumber\\
&=& 
\oint_{\partial\Sigma}(k^\alpha \Theta^\beta - k^\beta
\Theta^\alpha)\dSab 
\nonumber\\ 
&=& 
\int_\Sigma\na_\beta 
(k^\alpha \Theta^\beta - k^\beta \Theta^\alpha)\dSa 
\nonumber \\ 
&=& 
\int_\Sigma\na_\beta\Theta^\beta k^\alpha\dSa 
- \int_\Sigma \Lie_k \Theta^\alpha \dSa,   
\label{dq2_Aj}
\eeqn 
where we have used the relation $\nabla_\alpha k^\alpha =0$ 
to obtain the last equality.  
The integrand of the last term in Eq.~(\ref{dq2_Aj}) is written as
\beqn
\Lie_k \Theta^\alpha \,\dSa
&=& (\epsilon + p) q^\alpha{}\!_\beta\, \Lie_k \xi^\beta \,\dSa
\nonumber\\
&&
\,+\,A_\beta(j^\alpha \Lie_k \xi^\beta - j^\beta \Lie_k \xi^\alpha)
\,dS_\alpha
\nonumber\\
&=& (\epsilon + p) u^\alpha u_\beta\, \Lie_k \xi^\beta \,\dSa
\nonumber\\
&&
\,+\,A_\beta(j^\alpha \Lie_k \xi^\beta - j^\beta \Lie_k \xi^\alpha)
\,dS_\alpha, 
\label{eq:LieTh_Aj}
\eeqn
where we used the fact that $\xi^\alpha$ 
as well as its Lie derivative along $k^\alpha$ is spatial 
$\Lie_k \xi^\alpha \na_\alpha t = 0$. 
These two terms in Eq.~(\ref{eq:LieTh_Aj}) cancel out with 
the last terms of Eqs.~(\ref{eq:Dlhuv}) and (\ref{eq:DlAj}).
Note that the current $j^\alpha$ respects the symmetry 
$\Lie_k j^\alpha = 0$.

Finally, we obtain an expression for $\dl (Q-\sum\limits_i Q_i)$: 
\begin{widetext}
\beqn
&&\!\!\!\!\!\!\!\!\!\!\!\!\!\!\!\!\!\!\!
\delta \Big(Q-\sum_i Q_i\Big)  
\nonumber\\
&=& 
\int_\Sigma \left\{\,\frac{T}{u^t} \Dl s \, \rho u^\alpha\, \dSa
\right.
\,+\,\left[\,\frac{h}{u^t}+hu_\beta  v^\beta  \,\right]
\Dl(\rho u^\alpha\, \dSa)
\,+\,v^\beta \Dl(hu_\beta)\rho u^\alpha\,\dSa 
\nonumber \\ 
&&\left.\phantom{\frac12}\!\!\!\!
\,-\, A_\beta k^\beta\, \Dl (j^\alpha \dSa)
\,-\,(j^\alpha k^\beta - j^\beta k^\alpha )
\Dl A_\beta \,\dSa
\,\right\}
\,-\,
\frac1{4\pi}\, \dl\int_\Sigma  F^{\alpha\gamma} \Lie_k A_\gamma dS_\alpha 
\,+\,\frac1{4\pi}\, \dl\int_{\pa\Sigma}
k^\gamma A_\gamma F^{\alpha\beta} dS_{\alpha\beta}
\nonumber\\
&-& \frac{1}{8\pi}\,\dl \int_\Sigma 
\left[\, \Gab - 8\pi (\Tab + \TabF)\,\right] k^\beta dS_\alpha
\,-\,\frac1{4\pi}\,\dl \int_\Sigma k^\gamma A_\gamma 
(\na_\beta \Fabu - 4\pi j^\alpha) dS_\alpha
\nonumber \\
&+& \int_\Sigma \left\{\,
\frac1{16\pi}\left[\,\Gabu -8\pi(\Tabu+\TabFu)\,\right]
\dl \gabd \, 
\,+\, \frac1{4\pi}(\na_\beta\Fabu - 4\pi j^\alpha)
\dl A_\alpha 
\,+\, \xi^\alpha\left[\,\na_\beta\Tba - \Fabd j^\beta 
\,\right] 
\right\}k^\gamma\, dS_\gamma.  \quad
\label{eqApp:dlQ_Aj}
\eeqn
Note that $k^\alpha \dSa = \sqrt{-g}d^3 x$.
When the field equations, their perturbations, and equations 
of motion are satisfied, 
using $\Lie_k A_\alpha = 0$, and Eq.~(\ref{eq:epotconst}) 
noting $\int_{\pa\Sigma}=\oint_{S}-\sum_i\oint_{{\cal B}i}$, 
Eq.~(\ref{eqApp:dlQ_Aj}) is rewritten 
\beqn
\delta \Big(Q-\sum_i Q_i\Big)  
&=& 
\int_\Sigma \left\{\,\frac{T}{u^t} \Dl s \, \rho u^\alpha\, \dSa
\right.
\,+\,\left[\,\frac{h}{u^t}+hu_\beta  v^\beta  \,\right]
\Dl(\rho u^\alpha\, \dSa)
\,+\,v^\beta \Dl(hu_\beta)\rho u^\alpha\,\dSa 
\nonumber \\ 
&&\left.\phantom{\frac12}\!\!\!\!
\,-\, A_\beta k^\beta\, \Dl (j^\alpha \dSa)
\,-\,(j^\alpha k^\beta - j^\beta k^\alpha )
\Dl A_\beta \,\dSa
\,\right\}
\,-\,\sum_i\frac1{4\pi}\, \dl\oint_{{\cal B}i}
k^\gamma A_\gamma F^{\alpha\beta} dS_{\alpha\beta}.
\label{eqApp:dlQ}
\eeqn
\end{widetext}

\section{Calculation of Eq.~(\ref{eq:BO_LFterm})} \label{s:cal_BO_LFterm}

A relation used in Eq.~(\ref{eq:BO_LFterm}) is proved in 
\cite{BB06}, which is repeated here for a reference.  
Consider a closed 2-form $\Fabd$ ($(dF)_{\albega} = 0$), and 
a vector $N^\alpha$ such that 
$F_\albe N^\beta  = 0$.  
Then, for any vector $ q^\alpha$, a relation 
\beq
(d\eta)_\albe\,N^\beta\,=\, F_\albe\, \Lie_q N^\beta
\label{eq:lemma}
\eeq
is satisfied, where $\eta_\alpha$ is defined by 
$\eta_\alpha = F_\albe\,q^\beta$.  
This can be shown as follows:
\beqn
(d\eta)_\albe N^\beta 
&=& 
(d(F\cdot q))_\albe N^\beta 
\nonumber\\
&=&
\left[(q\cdot dF)_\albe \,-\, \Lie_q F_\albe\right] N^\beta 
\nonumber\\
&=&
F_\albe \Lie_q  N^\beta. 
\eeqn
The Cartan identity was used in the second equality 
and the relation $F_\albe N^\beta  = 0$ in third one.

\section{First integral of MHD-Euler equation in BGSM formulation}
\label{secApp:BGSM}

A formulation for uniformly rotating axisymmetric stars with 
poloidal magnetic fields is derived in \cite{BGSM93}.  In this 
section, we show that the Bekenstein and Oron formulation of ideal MHD 
 includes a first integral of the MHD-Euler 
equation derived in the BGSM formulation, assuming 
the same symmetry and suitably choosing an auxiliary
vector $q^\alpha$ in the current (\ref{eq:BOcur}).  

In the BGSM formulation, a stationary, axisymmetric and circular spacetime 
is assumed.  And more specifically the flow field of rotating star 
is assumed to be uniform; with a constant angular velocity $\Omega$, 
4-velocity is written 
$u^\alpha = u^t k^\alpha = u^t(t^\alpha + \Omega \phi^\alpha)$ 
where $t^\alpha$ and $\phi^\alpha$ are killing vectors.  

Carter has shown \cite{Carter73} that in stationary, axisymmetric 
and circular spacetime, the vector 
potential and the current are such that  
$A_\alpha = A_t \na_\alpha t + A_\phi \na_\alpha \phi$ and 
$j^\alpha = j^t t^\alpha + j^\phi \phi^\alpha$ respectively.  
Since the vector potential $A_\alpha$ is assumed to 
respect the symmetry $\Lie_k A_\alpha =0$, 
the ideal MHD condition $\Fabd u^\beta=0$ implies, for a corotating flow, 
\beq
\Fabd k^\beta = -\Lie_k A_\alpha + \na_\alpha(A_\beta k^\beta) 
= \na_\alpha(A_\beta k^\beta) =0, 
\label{eq:pMHDcorot}
\eeq
hence 
\beq
A_\alpha k^\alpha = A_t + \Omega A_\phi = \mbox{constant}.
\eeq
Using this relation, the vector potential is written 
\beq
A_\alpha \,=\, A_\phi(\na_\alpha \phi-\Omega \na_\alpha t). 
\eeq
Note that $\na_\alpha \phi-\Omega \na_\alpha t$ is 
orthogonal to the helical vector, 
$k^\alpha (\na_\alpha \phi-\Omega \na_\alpha t) =0$.

Rewriting the current as 
\beq
j^\alpha = j^t k^\alpha + J \phi^\alpha, 
\eeq
with $J = j^\alpha  (\na_\alpha \phi-\Omega \na_\alpha t)
= j^\phi -\Omega j^t$, the Lorenz force becomes 
\beq \label{e:Lor_for}
\frac1{\rho}\Fabd j^\beta \,=\, \frac{J}{\rho}\Fabd \phi^\beta 
\,=\, \frac{J}{\rho}
\left[\,-\Lie_\phi A_\alpha + \na_\alpha(A_\beta \phi^\beta) \,\right].
\eeq
Then, with the symmetry $\Lie_\phi A_\alpha =0$, 
the MHD-Euler equation (\ref{eq:MHD-Euler}) is written 
\beq
k^\beta ( d(h\underbar{u}))_\beal 
\,=\, \frac{J}{\rho u^t} \na_\alpha(A_\beta \phi^\beta),
\eeq
or using $k^\beta (d(h\underbar{u}))_\beal 
= - \na_\alpha(hu_\beta k^\beta)
= \na_\alpha(h/u^t)$, 
\beq
\na_\alpha\left(\frac{h}{u^t}\right) - \frac{J}{\rho u^t} \na_\alpha A_\phi 
\,=\, 0.
\label{eq:MHD-Euler_BGSM}
\eeq
Hence, an integrability condition of this equation is 
\beq
\frac{J}{\rho u^t} 
\,=\, f(A_\phi).
\eeq

Equation~(\ref{eq:MHD-Euler_BGSM}) and MHD-Euler equation for 
the comoving flow (\ref{eq:ME_corot}) 
with the current (\ref{eq:BOcur}) agree if the relation 
\beq
\Fabd \Lie_k q^\beta = - \frac{J}{\rho u^t} \na_\alpha A_\phi 
\eeq
is satisfied.  
For example, if the vector $q^\alpha$ satisfies 
\beq
\Lie_k q^\alpha 
\,=\, -\frac{J}{\rho u^t}\phi^\alpha 
\,=\, f(A_\phi)\phi^\alpha, 
\eeq
the Bekenstein and Oron formulation 
becomes the BGSM formulation [cf. Eq.~(\ref{e:Lor_for})].

\section{Calculation of $\dl (Q - \sum\limits_i Q_i)$ for the 
Lagrangian with Bekenstein and Oron's interaction term}
\label{secApp:dlQBO}

In the Bekenstein and Oron theory, 
the ideal MHD condition $\Fabd u^\beta = 0$ is imposed by 
adding a constraint to the Lagrangian density with 
a Lagrange multiplier $q^\alpha$, 
\beq
\LagI\,=\,\Fabd\, \rho u^\alpha q^\beta\sqrt{-g}.
\eeq
This term replaces an interaction term, $A_\alpha j^\alpha\sqrt{-g}$, 
of the field and electric current.  
The variation of  $\LagI$ becomes, 
\beqn
&&
\frac1{\sqrt{-g}}\dl\LagI
\,=\,
-\na_\beta\dl A_\alpha (\rho u^\alpha q^\beta-\rho u^\beta q^\alpha)
\nonumber\\
&&+\,\frac1{\sqrt{-g}}\Fabd
\left[
\Dl(\rho u^\alpha q^\beta \sqrt{-g})
\,-\,\Lie_\xi(\rho u^\alpha q^\beta \sqrt{-g})
\right], \qquad
\eeqn
The last term is calculated as 
\beqn
&-&\frac1{\sqrt{-g}}\Fabd\Lie_\xi(\rho u^\alpha q^\beta\sqrt{-g})
\nonumber\\
&=&
\rho u^\alpha q^\beta
\left[\,\xi^\gamma(dF)_{\gamma\albe}+d(\xi\cdot F)_\albe\,\right]
\nonumber\\
&+&
\na_\alpha(F_{\beta\gamma}\,\rho u^\gamma q^\beta\xi^\alpha), 
\eeqn
where the Cartan identity for the 2-form $\Fabd$, 
$\Lie_\xi \Fabd = \xi^\gamma (dF)_{\gamma\albe} +
(d(\xi\cdot F))_\albe$ is used, 
and 
\beqn
&&
\rho u^\alpha q^\beta (d(\xi\cdot F))_\albe
\,=\, (\rho u^\alpha q^\beta - \rho u^\beta q^\alpha)
\na_\alpha(\xi^\gamma F_{\gamma\beta})
\nonumber\\
&&=\, \xi^\alpha \Fabd \, j^\beta
\,+\,
\na_\alpha\left[\,(\rho u^\alpha q^\beta - \rho u^\beta q^\alpha)
\xi^\gamma F_{\gamma\beta}\,\right].
\eeqn
Hence, using $j^\alpha = \na_\beta(\rho u^\alpha q^\beta-\rho u^\beta q^\alpha)$, 
we have 
\beqn
&&
\frac1{\sqrt{-g}}\dl\LagI
\,=\,
\frac1{\sqrt{-g}}\Fabd
\Dl(\rho u^\alpha q^\beta \sqrt{-g})
\nonumber\\
&&
\,+\, j^\alpha \dl A_\alpha
\,+\, \xi^\alpha\left[\,
\Fabd\,j^\beta\,+\,\rho u^\beta q^\gamma(dF)_{\alpha\beta\gamma}
\,\right]
\,+\, \na_\alpha \ThIau, 
\nonumber\\
\eeqn
where 
\beqn
\ThIau 
&=& 
(\rho u^\alpha q^\beta - \rho u^\beta q^\alpha)
\dl A_\beta
\nonumber\\
&-&(\rho u^\alpha q^\beta \xi^\gamma
+ \rho u^\beta q^\gamma \xi^\alpha
+ \rho u^\gamma q^\alpha \xi^\beta)\, F_{\beta\gamma}.
\eeqn

To calculate the difference of Noether charge 
$\dl (Q - \sum\limits_i Q_i)$, we first associate 
$Q_K - \sum\limits_i Q_{Ki}$ with the Lagrangian 
(\ref{eq:LagBO}) as 
\beqn
&&
 Q_K -\sum_i Q_{Ki} 
\nonumber\\
&=&
-\, \int_\Sigma \Lag \,d^3x
\,+\,\int_\Sigma (\epsilon + p)u^\alpha u_\beta v^\beta dS_\alpha 
\nonumber\\
&&
\,-\,\frac1{4\pi}\int_\Sigma  F^{\alpha\gamma} \Lie_k A_\gamma dS_\alpha 
\,+\,\frac1{4\pi} \int_{\pa\Sigma}
k^\gamma A_\gamma F^{\alpha\beta} dS_{\alpha\beta}
\nonumber\\
&&
\,+\,\int_\Sigma \Fabd\,\rho u^\alpha q^\beta k^\gamma dS_\gamma 
\,-\, \int_\Sigma k^\gamma A_\gamma\,  j^\alpha dS_\alpha .
\nonumber\\
&&
\,-\, \frac{1}{8\pi} \int_\Sigma 
\left[\, \Gab \,-\, 8\pi (\Tab + \TabM)\,\right] k^\beta dS_\alpha
\nonumber\\
&&
-\,\frac1{4\pi} \int_\Sigma k^\gamma A_\gamma 
(\na_\beta \Fabu - 4\pi j^\alpha) dS_\alpha, 
\label{eq:qkBO}
\eeqn
which corresponds to Eq.~(\ref{eq:QK_Aj}).
The integrand of the fifth term in the r.h.s.~of Eq.~(\ref{eq:qkBO}) 
is rewritten 
\beqn
&& 
\Fabd\,\rho u^\alpha q^\beta k^\gamma dS_\gamma 
\,=\,
\Fabd\,\rho (k^\alpha + v^\alpha) q^\beta u^\gamma dS_\gamma 
\nonumber\\
&=& k^\alpha\Fabd\, q^\beta \rho u^\gamma dS_\gamma 
\,+\, v^\alpha \eta_\alpha\, \rho u^\gamma dS_\gamma,  
\eeqn
and combined with the sixth term as 
\beqn
&& v^\alpha \eta_\alpha\, \rho u^\beta dS_\beta
\,-\,\na_\alpha(k^\beta A_\beta)
\, q^\alpha \rho u^\gamma dS_\gamma 
\,-\, k^\beta A_\beta\,  j^\alpha dS_\alpha 
\nonumber
\\
&&\,=\,
 v^\alpha \eta_\alpha\, \rho u^\beta dS_\beta
\,-\, \Lie_q(k^\beta A_\beta \, \rho u^\gamma dS_\gamma) 
\label{eq:qkcurOB}
\eeqn
where Eq.~(\ref{eq:cur2}) and $\Lie_k A_\alpha=0$ were used.  
The integral of the last term of Eq.~(\ref{eq:qkcurOB}) over 
$\Sigma$ is rewritten a surface integral over $\pa\Sigma$
that vanishes, because of the gauge invariance under 
the transformation $q^\alpha \rightarrow q^\alpha + \lambda u^\alpha$ 
which can always be used to make 
$q^\alpha$ spatial, $q^\alpha \na_\alpha t = 0$.

The third line of Eq.~(\ref{eq:qkBO}) is replaced by 
$v^\alpha \eta_\alpha\, \rho u^\beta dS_\beta$, then 
a variation of the charge is calculated.  

A difference from the calculation of 
$\dl (Q_K - \sum\limits_i Q_{Ki})$ 
in Appendix \ref{secApp:dlQ} is the terms, 
\beqn
&&
\Dl\left[(\epsilon + p) u^\alpha u_\beta v^\beta \dSa\right] 
\,+\,\Dl(v^\beta \eta_\beta\, \rho u^\alpha \dSa )
\nonumber\\
&=& 
 (hu_\beta + \eta_\beta) v^\beta \Dl(\rho u^\alpha\dSa)  
\,+\, v^\beta \Dl(hu_\beta + \eta_\beta)\rho u^\alpha\dSa 
\nonumber\\
&+&
(\epsilon + p) u^\alpha u_\beta\, \Lie_k \xi^\beta \dSa 
\,+\, \Lie_k \xi^\beta \eta_\beta\, \rho u^\alpha \dSa, 
\label{eq:Dlepp}
\eeqn 
where $\Dl v^\beta = -\Dl k^\beta = \Lie_k \xi^\beta$ is used.  
In the calculation of $\dl (Q_L - \sum\limits_i Q_{Li})$, 
a term $\Lie_k \Theta^\alpha \,\dSa$ becomes, 
\beqn 
\Lie_k \Theta^\alpha \,\dSa
&=& (\epsilon + p) u^\alpha u_\beta\, \Lie_k \xi^\beta \,\dSa
\nonumber\\
&+&
\Lie_k \xi^\gamma F_{\gamma\beta}
(\rho u^\alpha q^\beta - \rho u^\beta q^\alpha) \,\dSa
\nonumber\\
&+&
(\dl A_\beta + \xi^\gamma F_{\gamma\beta})
(\rho u^\alpha \Lie_k q^\beta - \rho u^\beta\Lie_k q^\alpha )\,\dSa
\nonumber\\
\label{eq:LieThBO}
\eeqn 
where $\xi^\alpha$ and $\Lie_k \xi^\alpha$ are both spatial. 
The first term and a part of the second term in the r.h.s. of 
Eq.~(\ref{eq:LieThBO}) cancel out with the last two terms 
in Eq.~(\ref{eq:Dlepp}).  
With the Cartan identity, $\xi^\gamma F_{\gamma\beta}
=\Lie_\xi A_\beta - \na_\beta(\xi^\gamma A_\gamma)$
the last term of Eq.~(\ref{eq:LieThBO}) 
becomes 
\beqn
&&
(\dl A_\beta + \xi^\gamma F_{\gamma\beta})
(\rho u^\alpha \Lie_k q^\beta - \rho u^\beta\Lie_k q^\alpha )\,\dSa
\nonumber\\
&=& 
\left[\,\Dl A_\beta - \na_\beta(\xi^\gamma A_\gamma)\,\right]
(\rho u^\alpha \Lie_k q^\beta - \rho u^\beta\Lie_k q^\alpha )\,\dSa
\nonumber\\
&=& 
\Dl A_\beta 
\,(\rho u^\alpha \Lie_k q^\beta - \rho u^\beta\Lie_k q^\alpha )\,\dSa
\,+\,\xi^\gamma A_\gamma \Lie_k j^\alpha \,\dSa
\nonumber\\
&-&
\na_\beta\left[\,\xi^\gamma A_\gamma
 (\rho u^\alpha \Lie_k q^\beta -  \rho u^\beta \Lie_k q^\alpha)\,\right]\dSa, 
\eeqn
where the second term of the r.h.s.~of the last equality vanishes 
for the symmetry, $\Lie_k j^\alpha =0$, and an integral of the last term 
over $\Sigma$ vanishes for the Stokes theorem.  

Finally, we obtain an expression for $\dl (Q-\sum\limits_i Q_i)$
for the Bekenstein and Oron ideal MHD theory: 
\begin{widetext}
\beqn
&&\!\!\!\!\!\!
\delta (Q-\sum_i Q_i)  
\nonumber\\
&=& 
\int_\Sigma \left\{\,\frac{T}{u^t} \Dl s \, \rho u^\alpha\, \dSa
\,+\,\left[\,\frac{h}{u^t}+(hu_\beta + \eta_\beta) v^\beta  \,\right]
\Dl(\rho u^\alpha\, \dSa)
\,+\,v^\beta \Dl(hu_\beta + \eta_\beta)\rho u^\alpha\,\dSa 
\right.
\nonumber \\ 
&&\!\!\!\!\left.
\,-\,(\rho u^\alpha \Lie_k q^\beta - \rho u^\beta\Lie_k q^\alpha )
\Dl A_\beta \,\dSa
\,\right\}
\,+\, \frac1{4\pi}\,\dl \oint_{\pa\Sigma}
k^\gamma A_\gamma F^{\alpha\beta} dS_{\alpha\beta}
\,+\,
\int_\Sigma 
\Fabd u^\beta \left[\frac1{u^t} \Dl (q^\alpha \rho u^\gamma\, dS_\gamma)
\,+\, \Lie_k \xi^\alpha \rho \, q^\gamma \,dS_\gamma
\right]
\nonumber\\
\nonumber\\
&-& \frac{1}{8\pi}\,\dl \int_\Sigma 
\left[\, \Gab - 8\pi (\Tab + \TabM)\,\right] k^\beta dS_\alpha 
\,-\, \frac1{4\pi}\,\dl \int_\Sigma k^\gamma A_\gamma 
(\na_\beta \Fabu - 4\pi j^\alpha) dS_\alpha
\nonumber \\
&+& \int_\Sigma \left\{\,
\frac1{16\pi}\left[\,\Gabu -8\pi(\Tabu+\TabMu)\,\right]
\dl \gabd \, 
\,+\, \frac1{4\pi}(\na_\beta\Fabu - 4\pi j^\alpha)
\dl A_\alpha 
\right.
\nonumber \\
&&\left.\phantom{\frac12}
+\ \xi^\alpha\left[\,\na_\beta\Tba - \Fabd j^\beta 
 - \rho u^\beta q^\gamma(dF)_{\alpha\beta\gamma}\,\right] 
\!\!\!\!\!\phantom{\frac12}\right\}k^\delta\, dS_\delta.  
\label{eqApp:dlQ_AjBO}
\eeqn
\end{widetext}
This expression is compared with Eq.~(\ref{eqApp:dlQ_Aj}).  
Note that, in the second line of Eq.~(\ref{eqApp:dlQ_AjBO}), 
the circulation of magnetized flow explicitly appears 
as in Eq.~(\ref{eq:1stlaw_BBcur}).

\end{document}